\documentclass[11pt]{article}

\RequirePackage{amsthm,amsmath,amsfonts,amssymb}
\RequirePackage[authoryear]{natbib}
\usepackage[colorlinks=true, citecolor=blue, linkcolor=blue, urlcolor=blue]{hyperref}
\usepackage{doi}
\RequirePackage{graphicx}
\usepackage{subcaption}
\usepackage{caption} 
\usepackage{appendix}
\usepackage{booktabs} 
\usepackage{dsfont}
\usepackage{float}  
\usepackage{lineno}
\usepackage{multirow}
\setcitestyle{round}
\usepackage[margin=1.1in]{geometry}
\date{}
\usepackage{fancyhdr}
\fancypagestyle{firstpage}{%
  \fancyhf{}
  \rhead{\thepage}
  
}
\begin{document}
\title{Estimating the distance at which narwhal respond to disturbance: a penalised threshold hidden Markov model}
\maketitle
\vspace{-2.5em}
\begin{center}
{\normalsize
Fanny Dupont$^{1}$,
Marianne Marcoux$^{2}$,
Nigel E. Hussey$^{3}$,
Jackie Dawson$^{4}$,
Marie Auger-M\'{e}th\'{e}$^{1,5}$
}\\[0.3em]
{\footnotesize
$^{1}$Department of Statistics, University of British Columbia \\
$^{2}$Freshwater Institute, Fisheries and Oceans Canada\\
$^{3}$Department of Integrative Biology, University of Windsor \\
$^{4}$Department of Geography, University of Ottawa\\
$^{5}$Institute for the Oceans and Fisheries, University of British Columbia\\[0.3em]
\textit{Correspondence:} fanny.dupont@stat.ubc.ca, auger-methe@stat.ubc.ca
}
\end{center}

\begin{abstract}
Understanding behavioural responses to disturbance is vital for wildlife conservation. In the Arctic, declining sea ice has opened new shipping routes, increasing the need to quantify the distance at which marine mammals react to vessel presence, which can guide mitigation policies such as vessel slow-down regulations and avoidance areas. Using telemetry data to identify distances linked to deviations from normal behaviour requires advanced statistical models, such as threshold hidden Markov models (THMMs). While THMMs are powerful, they do not assess whether an estimated threshold reflects a distinct behavioural shift. We introduce a lasso-penalised THMM built on computationally efficient methods, along with a new penalised quasi-restricted maximum-likelihood estimator. Our framework estimates thresholds and assesses whether disturbance effects are distinguishable from baseline behaviour. Simulations show that the lasso penalty effectively shrinks spurious threshold effects towards zero. Applied to narwhal \textit{(Monodon monoceros)} movement data, our analysis suggests that narwhal react to vessels up to 3.4 kilometres away, decreasing movement persistence and diving deeper. Overall, we provide a broadly applicable framework for quantifying behavioural responses to stimuli, with applications ranging from disturbance thresholds to estimating detection distances, such as how far elephants can detect water. We also provide a tutorial and open-source implementation.
\end{abstract}

\vskip1cm
\section{Introduction}
Understanding changes in animal behaviour in response to human disturbance is crucial for mitigating the effects of anthropogenic activity on ecosystems. In marine environments, this disturbance often manifests as underwater noise, a recognised stressor to marine life (\citealp{southall_marine_2019,pame2019underwater}). Human activity can create underwater noise through both intentional (e.g., sonar, military exercises and airguns used for seismic exploration) and incidental (e.g., propeller cavitation, engine noise) sources, with growing evidence that marine mammals modify their behaviour in response  (narwhal (\textit{Monodon monoceros}) in \citealp{tervo_narwhals_2021,tervo_stuck_2023}; marine tucuxis (\textit{Sotalia fluviatilis}) in \citealp{carrera_response_2008}; belugas (\textit{Delphinapterus leucas}) in \citealp{martin_exposure_2023}; and beaked whales (\textit{Ziphius cavirostris}) in \citealp{michelot_continuous-time_2022}). The impacts of underwater noise are especially relevant in the Arctic, where shipping activity is expanding into areas that had remained acoustically undisturbed until recently (\citealp{pizzolato_changing_2014,pizzolato_influence_2016}). 

Western science studies (\citealp{tervo_narwhals_2021}; \citealp{tervo_stuck_2023, delporte_varying_2025}) and Inuit knowledge (\citealp{aariak2019qikiqtani}) report that narwhal modify their diving behaviour when exposed to anthropogenic noise, but their responses to human disturbance are complex, and many studies document contrasting reactions. For example, \cite{tervo_stuck_2023} found that narwhal reduce deep foraging dives and increase shallow diving in response to ship noise and airgun pulses. Using the same dataset, \cite{delporte_varying_2025} found a decrease in movement persistence and an increase in swimming speed. In contrast, \cite{golder_maryriver_2020} found that narwhal decrease their time at the surface when exposed to vessel noise and \cite{williams2017paradoxical} reported deep “escape dives” following entanglement or acute stress. Given that these marine mammals hold great cultural value to Inuit communities (\citealp{reeves1992recent}) and are considered to be among the most vulnerable Arctic species to climate change (\citealp{laidre2008quantifying,nirb2021maryriver}), it is critical to improve our understanding of narwhal responses to increasing shipping activities in the Arctic. Specifically, resolving how narwhal react to increasing vessel traffic will help inform mitigation strategies. 


Telemetry data are commonly used to study animal behaviour (e.g., \citealp{hussey2015aquatic, kays2015terrestrial}), as they provide fine-scale information that is often difficult to obtain through direct observation. However, most methods used to identify marine mammal response to disturbances using telemetry data have been developed in the context of controlled exposure experiments (e.g., \citealp{deruiter_multivariate_2017,isojunno_individual_2017, tervo_narwhals_2021, michelot_continuous-time_2022,tervo_stuck_2023,delporte_varying_2025}). Controlled experiments are well suited for studying disturbance effects, as they provide clear pre- and during-disturbance periods that enable direct behavioural comparisons. In practice, however, such ideal conditions are rarely available, and disturbance is difficult to pinpoint because behavioural responses often change gradually with increasing exposure. In that case, researchers are often limited to individual-based inference, analysing each animal separately to infer periods of disturbance (e.g., from dive profiles; \citealp{mikkelsen2019long}). This approach is inherently limited and complicates the process of defining a disturbance threshold applicable to an entire species. Consequently, there is a need for the development of a standardised method to infer disturbance responses at the population or species level from telemetry data collected outside controlled experiments. 

Threshold effect models are particularly well-suited for addressing these challenges, as they can detect sudden, significant shifts in complex time-series data. They have been widely used across diverse fields, such as mathematical finance (\citealp{mohsin_s_khan_and_abdelhak_s_ssnhadji_threshold_2001}), medicine (\citealp{fong_chngpt_2017}), ecology (\citealp{scheffer_catastrophic_2001}) and oil price trends analysis (\citealp{zhu_hidden_2017}). In disturbance modelling, two regimes are considered: (1) baseline and (2) disturbed. 
A key goal of threshold effect models is to determine the critical level of a covariate (e.g., temperature, distance to vessels, or drug dosage) beyond which the time series undergoes a regime shift. Therefore, threshold effect analyses require an understanding of the baseline process of the time series to accurately measure deviations from the baseline during disturbance. 

Threshold hidden Markov models (THMMs; \citealp{zhu_hidden_2017}) are hidden Markov models (HMMs) with a two-component mixture in their transition probability matrix. Hidden Markov models assume that the observed sequence is generated by an underlying Markov process over a finite set of hidden states. The states carry information about the phenomenon of interest (survival status, \citealp{mcclintock_uncovering_2020}; purchase types of a consumer, \citealp{srivastava_credit_2008}; phonemes in speech, \citealp{rabiner_introduction_1986}). In the context of animal telemetry data, they are usually interpreted as proxies for animal behaviour (e.g., resting, foraging; \citealp{morales_extracting_2004,pohle_selecting_2017}). Threshold HMMs further allow the latent Markov chain to switch between two distinct regimes, thereby separating transitions into a baseline and a disturbed dynamical process. 
The mixture probability governing the switch between the two regimes is controlled by a step function ($0$ or $1$) that activates when a covariate exceeds a certain threshold. By combining regime-specific transitions with the threshold-driven mixture probability, THMMs can quantify deviations from a baseline regime in response to disturbances and, crucially, estimate the threshold that triggers regime shifts. 

While THMMs are powerful, they suffer from two important limitations. First, their practical implementation requires computationally intensive grid searches (i.e., fitting a separate model for each candidate threshold value), which becomes prohibitive for high-dimensional or large datasets like animal movement data (\citealp{zhu_hidden_2017,patterson_statistical_2017}). Second, while THMMs allow reliable threshold estimation (\citealp{zhu_hidden_2017}), there is no method to assess whether the disturbed component reflects a distinct shift in behaviour. As a mixture model, the absence of disturbance in a THMM implies that the baseline and disturbed components are equivalent, and the estimated threshold has no meaningful interpretation. As such, we require a computationally-efficient method to estimate thresholds in THMMs along with a principled method to ensure that these estimates correspond to measurable behavioural changes.

Assessing whether the disturbed component improves the model is akin to component selection in mixture models, a notoriously challenging task. While methods like the likelihood ratio test (LRT) are popular for their simplicity, they lack valid asymptotic distributions in finite mixtures (\citealp{gassiat_likelihood_2000,lo_likelihood_2005}). Modified LRT for homogeneity testing offers solutions for standard mixture models (\citealp{chen_modified_2001}), but these are not directly applicable to THMMs. While bootstrap likelihood ratio tests (BLRTs) are adequate for mixture models (\citealp{mclachlan_bootstrapping_1987,feng1996using,lo_likelihood_2005}), their application in THMMs remains unexplored. Moreover, BLRTs are expensive to fit, which can be prohibitive in animal telemetry data analysis (\citealp{patterson_statistical_2017}). Penalised likelihood methods provide an elegant and effective framework for order selection in mixture models and HMMs (\citealp{chen_order_2008,hung_hidden_2013,lin_order_2022, dupont2025improved}). Thus, we propose a penalised likelihood approach for the related task of identifying a disturbed component from animal telemetry data.

We present a computationally efficient approach for estimating threshold parameters in THMMs using a lasso penalty to shrink spurious estimates towards zero (\citealp{kang2019bayesian}; \citealp{yao2020latent}). 
For computational efficiency, we build on methods developed for generalised linear mixed models (GLMMs) by \cite{laird1982random} and adapted to HMMs with penalised splines by \cite{koslik_efficient_2024}. Specifically, we interpret the lasso penalty as a distributional prior on random threshold effects. We then approximate the corresponding marginal likelihood using the Laplace approximation, integrated within a quasi-restricted maximum likelihood (qREML) framework, to reduce the computational overhead associated with selecting the penalty strength. Computational efficiency is further improved via the use of a smooth logistic function to approximate the step function instead of using grid search (\citealp{fong_chngpt_2017}). We evaluate our approach through a simulation study covering a range of threshold scenarios and sample sizes, and by applying it to narwhal movement data. In our case study, the threshold function depends on vessel proximity, allowing us to estimate a single vessel disturbance threshold in kilometres that is shared across all tracks. To our knowledge, this constitutes the first model-based estimate of a disturbance threshold in movement ecology, with direct relevance for informing mitigation policies. More broadly, our approach represents a crucial step towards establishing THMMs as a practical tool for applied time-series analysis. We provide a tutorial and R source code on GitHub (\url{https://github.com/Fanny-Dupont/THMM}) for practitioners to apply this method to their own datasets.

\section{Narwhal and vessel data}
\subsection{Narwhal movement data} 
Our analysis focuses on the Qikiqtaaluk (Baffin) region in Nunavut, Canada. During the summer of 2017, 18 narwhal were equipped with electronic tags in Tremblay Sound ($72^{\circ}21.389\text{'N}, -81^{\circ}05.855\text{'W}$). All capture and tagging protocols were approved by the Fisheries and Oceans Animal Care Committee and a Licence for Scientific Purposes was granted (permit \text{\#AUP 40, S-17/18-1017-NU}). We obtained FastLoc GPS data (August–October 14) for $11$ narwhal (five females and six males). Satellite tags also included time-depth recorders sampling at 75-second intervals. To minimise handling effects, we excluded the first 24 hours of post-capture data (\citealp{shuert_assessing_2021,shuert_decadal_2022}). Similar to \cite{auger-methe_including_2025} and \cite{shuert_putting_2025}, location data (i.e., latitude and longitude) were corrected for error by fitting a continuous-time correlated random walk with the \texttt{R} package \texttt{crawl} (\citealp{johnson2018crawl}), with a resolution of one location every $30$ minutes (\citealp{johnson2008continuous}). Tracks with gaps larger than 90 minutes were split and assigned new independent IDs, and only those containing at least 10 points were kept (\citealp{storrie2023beluga}). This procedure resulted in a total of $8{,}603$ location points across $231$ tracks. The distance to shore was recorded for each whale location, as this covariate has been shown to be important for explaining narwhal behaviour (\citealp{heide-jorgensen_behavioral_2021, dupont2025improved, hornby2025behavioural}).

We then converted longitude and latitude into two data streams (\citealp{morales_extracting_2004}): step length (distance between consecutive locations) and turning angle
(change in bearing between consecutive steps). Raw dive data were processed to extract maximum depth within 30-minute intervals. Given narwhal’s specialisation in deep diving, maximum dive depth is a key indicator for detecting behavioural modifications caused by nearby vessels (\citealp{williams2017paradoxical}; \citealp{tervo_stuck_2023}). \vskip0.5cm
\subsection{Vessel Automatic Identification System} 
\begin{figure}
    \centering
    \includegraphics[width=10.5cm,height=8cm]{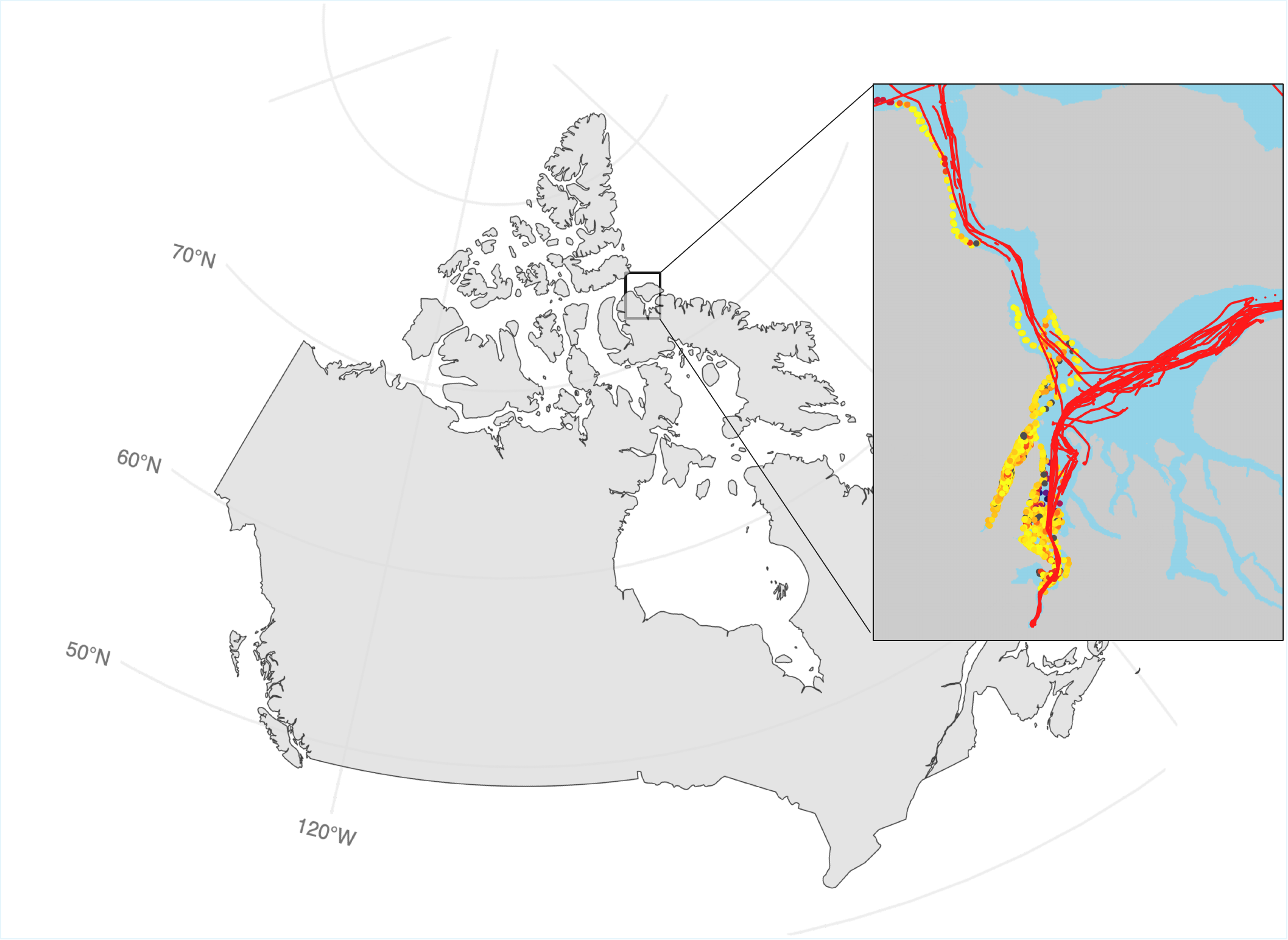}
    \caption{Narwhal locations at 30-minute resolution, shown as coloured points ranging from yellow (shallow) to dark purple/black (deep), and vessel tracks, shown as red lines, for the first week of August 2017.}
    \label{tracksdata}
\end{figure}

To determine whether vessel presence affects narwhal behaviour, we use vessel geographic coordinates collected in 2017 via the satellite Automatic Identification System (AIS; exactEarth, Cambridge, ON). Only vessels longer than $20$m are required to carry an AIS responder, resulting in less data from smaller vessels (\citealp{CanadaNavigationSafety2020}). Vessel data were corrected for error and missing data by fitting a continuous-time correlated random walk with a resolution of one location every minute with \texttt{crawl}. For each narwhal location, we calculated the geodesic distance to all vessels within a $\pm$30-second time window and recorded the presence of land (e.g., islands) intersecting the direct path, as this can block or reduce sound transmission. Following \cite{tervo_stuck_2023} and \cite{delporte_varying_2025}, we quantified vessel exposure as the inverse distance (km$^{-1}$) between the whale and nearest vessel. Figure~\ref{tracksdata} shows the narwhal tracks along with the AIS tracks for the first week of August 2017.


\section{Modelling disturbance reaction with THMMs}
\subsection{Method}
Previous studies have shown that vessel noise can alter narwhal behaviour, such as reducing the probability of initiating deep dives (\citealp{tervo_narwhals_2021}; \citealp{tervo_stuck_2023}). Threshold HMMs naturally accommodate such behavioural changes by modelling the hidden state process as a mixture of two regimes: (a) a baseline regime, representing behaviour in the absence of disturbance and (b) a disturbed regime, capturing behavioural responses to disturbance. Each regime is governed by a transition probability matrix (TPM). 
This structure makes THMMs well suited for studying responses to disturbance, as it allows us to quantify how exposure to vessels alters the probability of switching between existing behaviours without assuming that entirely new behaviours emerge.


\subsubsection{Model description}
\label{modeldescription}

Using a THMM, we investigate the effect of a univariate time series $\{u_t\}_{t=1}^T  = (u_1, \ldots, u_T)$ of length $T$, with $u_t \geq 0$ for all $t$, for example the exposure to the closest vessel, on a sequence of (potentially multivariate) observations $(\boldsymbol{Y}_1, \ldots, \boldsymbol{Y}_T)$, such as step length or maximum dive depth of narwhal recorded at regular time intervals. Threshold HMMs involve a sequence of discrete hidden states $(S_1,\ldots,S_T)$ that form a Markov chain and take values from $1$ to $N$. The model distinguishes between the baseline regime $B$ and the disturbed regime $D$. The distribution of the state process is fully determined by an initial state distribution and a TPM, with one such pair for each regime. Thus, conditional on being in regime $k \in \{B,D\}$, the initial state $S_1$ is distributed according to the row vector $\boldsymbol{\delta}^{(k)} = (\delta^{(k)}_1,\ldots,\delta^{(k)}_N)$, and the state transitions follow the regime-specific TPM $\boldsymbol{\Gamma}_t^{(k)}$, with $\Gamma_{t,ij}^{(k)} = \mathbb{P}^{(k)}(S_{t+1} =j| S_t = i)$. 
We include covariate effects in the transition probabilities via a multinomial logit link as follows (\citealp{mcclintock_momentuhmm_2018}):
\begin{equation}
\Gamma^{(k)}_{t,ij} = \frac{e^{c^{(k)}_{t,ij}}}{\underset{l}{\sum} e^{c^{(k)}_{t,il}}}, \text{ for } k \in \{B,D\}, \quad
c^{(k)}_{t,ij} = 
\begin{cases}
\alpha_{ij}^{k,0} + \underset{m=1}{\overset{C}{\sum}} \alpha_{ij}^{k,m} \omega_t^m, & \text{for } i \neq j \\
0, & \text{for } i = j,
\end{cases}
\end{equation}
where $(\omega_{t}^1,\ldots,\omega_{t}^C)$ is a vector of $C$ explanatory variables at time $t$ and $\boldsymbol{\alpha}^{(k)} = (\boldsymbol{\alpha}^{(k)}_{ij})_{i,j=1,\ldots,N^{}}$ denotes the collection of regression coefficient vectors for all state transitions of regime $k$. The observations $(\boldsymbol{Y}_1, \ldots, \boldsymbol{Y}_T)$ are assumed independent given the states. The current state $S_t = i$ determines the state-dependent density $f(\cdot \mid \gamma _i)$ that generates the observation $\boldsymbol{Y}_t$, where $\gamma_i$ is a state-dependent parameter describing the state-dependent
distribution. The mixture probability $\nu_{\beta_0}(u_t) $ for transitioning between regimes $B$ and $D$ at time $t$, is defined as follows:

\begin{equation}
\label{nu}
\nu_{\beta_0}(u_t)
=
\left\{
\begin{aligned}
\mathds{1}_{u_t > {1}/{\beta_0}}, &\quad \beta_0 > 0, \\
0, &\quad \beta_0 = 0 .
\end{aligned}
\right.
\end{equation}
Note that this is equivalent to $\nu_{\beta_0}(u_t)=\mathds{1}_{\beta_0u_t > {1}},$ since, for $\beta_0 = 0$, $\mathds{1}_{0 > {1}}=0$. When $\beta_0\neq0,$ we call $1/\beta_0$ the threshold.
Observe that equation \eqref{nu} only covers settings in which larger values of $u_t$ correspond to increased disturbance. This formulation is appropriate for covariates such as exposure to the nearest vessel. If one wishes to model the opposite relationship instead (i.e., lower values of $u_t$ correspond to increased disturbance), the covariate can be transformed accordingly (e.g., by using $1/u_t$). The likelihood function of a THMM for model parameters $(\boldsymbol{\theta}, \beta_0)$, where $\boldsymbol{\theta}= ( \boldsymbol{\delta}^{(B)}, \boldsymbol{\delta}^{(D)},\boldsymbol{\gamma}_1,\ldots,\boldsymbol{\gamma}_N,\boldsymbol{\alpha}^{(B)},\boldsymbol{\alpha}^{(D)})$, can be written as follows:

\begin{equation}
\label{mixtboats}
\begin{aligned}
\mathcal{L}(\boldsymbol{\theta}, {\beta}_{0}) = 
\big[
    \big(1-\nu_{{\beta}_{0}}({u}_{1})\big)  
    {\boldsymbol{\delta}^{(B)}}^{\top} \mathbf{P}(\boldsymbol{Y}_{1})
    +
    \nu_{{\beta}_{0}}({u}_{1})
 {\boldsymbol{\delta}^{(D)}}^{\top} \mathbf{P}(\boldsymbol{Y}_{1})
\big]\\
{\big (}\underset{t=2}{\overset{T}\prod}[1-\nu_{{\beta}_{0}}({u}_{t})]\boldsymbol{\Gamma}_t^{(B)}\textbf{P}(\boldsymbol{Y}_{{t}}) + 
\nu_{{\beta}_{0}}({u}_{t})\boldsymbol{\Gamma}_t^{(D)}\textbf{P}(\boldsymbol{Y}_{{t}}){\big )}\boldsymbol{1},
\end{aligned}
\end{equation}
where the $N^{} \times N^{}$ matrix $ \textbf{P}(\boldsymbol{Y}_{t})$ is diagonal with $(i,i)^{th}$ entry $f(\boldsymbol{Y}_{t}| \gamma_i)$. We use the notation $\boldsymbol{1}$ for the column vector of $N$ ones. The value of ${\beta}_0$ governs the presence of the second (disturbed) regime through $\nu_{{\beta}_0}(\cdot)$. Transitions between regimes can occur only when the covariate ${u_t}$ crosses the threshold $1/\beta_0$, that is, when $u_t$ moves from below to above the threshold or vice versa. Thus at each time $t$, the time series can transition from one regime to the other. 

We adopt a frequentist framework and obtain maximum likelihood estimates $(\boldsymbol{\hat{\theta}},\hat{\beta}_0)$, as is standard in ecological applications of HMMs (further details on the implementation are provided in the Supplementary Material; \citealp{mcclintock_momentuhmm_2018, mcclintock_worth_2021}). As a discontinuous step function, $\nu_{\beta_0}(\cdot)$ is unsuitable for gradient-based optimisation. We approximate it using a smooth, two-parameter logistic function as follows: 
\begin{equation}\label{twoparamlogistic}\nu_{\beta_0}(u_t)\approx\left[1 + \exp^{-b\left({\beta}_0{u}_t - 1\right)}\right]^{-1},\end{equation}
where $b$ controls the sharpness of the transition (\citealp{fong_chngpt_2017}). Since the effect of $b$ depends on the scale of the covariate, we standardise $\{u_t\}_{t=1}^T$ to the interval $[0, 1]$ to ensure a consistent approximation across applications.  
In practice, the covariate is standardised to the unit interval prior to model fitting, so the unstandardised covariate $\{u_t\}_{t=1}^T$ may take values in $\mathbb{R}$.
If $\hat{\beta}_0$ denotes the estimated parameter on the standardised scale, the implied threshold on the original scale of the covariate is given by
\begin{equation}
\frac{1}{\hat{\beta}_0}
\bigl(\max_{t} u_t - \min_{t} u_t\bigr)
+ \min_{t} u_t.
\end{equation}
Hereafter, the notation $\{u_t\}_{t=1}^T$ refers to the standardised covariate. The THMM described by model~(\ref{mixtboats}) can then capture the effect of covariates at two levels by identifying (a) the threshold value of $u_t$ that induces a shift in behaviour, and (b) the nature and extent of the resulting disturbance.

Model (\ref{mixtboats}) can be extended to accommodate a multivariate covariate $\boldsymbol{u}_t = (u_{1,t}, \ldots, u_{p_2,t})^\top$ by using $\boldsymbol{\beta}_0 = (\beta_{0}^1, \ldots, \beta_{0}^{p_2})^\top$ and $\nu_{\boldsymbol{\beta}_0}(\boldsymbol{u}_t) = \mathds{1}_{(\boldsymbol{\beta}_0^\top \boldsymbol{u}_t > 1)}$, which equals one if $\boldsymbol{\beta}_0^\top \boldsymbol{u}_t > 1$ and zero otherwise. 
For example, the interaction between the exposure to the closest vessel and the presence of land between the vessel and the whale can be represented using two covariates: $u_{1,t}$, defined as the {standardised} exposure to the closest vessel when land lies between the vessel and the whale (and $0$ otherwise), and $u_{2,t}$, defined as the {standardised} exposure to the closest vessel when no land lies between them (and $0$ otherwise). Since these categories are mutually exclusive (a vessel cannot be separated from the whale by land and not separated by land at the same time), the positivity constraint can be applied to each element of $\boldsymbol{\beta}_0$ separately and we can estimate land-specific exposure thresholds for disturbance. The ability to estimate covariate-specific thresholds is highly advantageous, as it allows us to assess whether the presence of land between a vessel and a whale attenuates vessel noise and consequently reduces the disturbance threshold. In contrast, when the covariates are continuous (e.g., $u_{1,t}$ is the \textit{exposure} from the closest vessel and $u_{2,t}$ its \textit{speed} at time $t$), the mixture probability depends on all covariates jointly. This joint dependence complicates estimation, especially in regions where multiple covariates approach their decision boundary simultaneously. In such cases, the model may have trouble separating the individual effects of each covariate, which can cause identifiability issues and lead to potential bias in the estimated thresholds. Thus, we do not explore this case further. From this point onward, multivariate covariates refer to the mutually exclusive structure described above.

\subsubsection{Penalised likelihood estimation}
\label{ple}

The null (i.e., no disturbance, one regime) and alternative (two regimes: baseline and disturbed) models are nested. The null model is a special case of the alternative either when $\Gamma_t^{(B)} = \Gamma_t^{(D)}, $ for all $t\geq 0$ (making the threshold irrelevant) or when $\nu_{\boldsymbol{\beta}_{0}}(\cdot) \equiv 0$ or $1$ (effectively using only one TPM). Thus, the model suffers from a lack of identifiability due to the non-unique representation of the null hypothesis in the alternative model's parameter space. 
Consequently, standard asymptotic results for nested models do not apply, and conventional asymptotic theory cannot be used for constructing confidence intervals.

Since LRTs are invalid for component selection in mixture models and BLRT methods are computationally intensive (\citealp{mclachlan_bootstrapping_1987,gassiat_likelihood_2000,mclachlan2000finite,lo_likelihood_2005,dziak_effect_2014}), we propose a computationally efficient method using a lasso-penalised likelihood combined with a qREML approach to estimate parameters and select the best model. The inclusion of the disturbed component is governed by lasso regularisation applied to $\boldsymbol{\beta}_0$. In the univariate case, a single $\beta_0$ is estimated. In the multivariate case with mutually exclusive covariates, a separate threshold is estimated for each covariate, independently of the others. A disturbance effect is excluded whenever the element of $\boldsymbol{\beta}_0$ associated with a given covariate is shrunk towards zero by the lasso penalty. Thus, if all elements of $\boldsymbol{\beta}_0$ are shrunk towards zero, the model reduces to a single-component THMM; if only a subset are nonzero, only the corresponding covariates induce a disturbed component. To our knowledge, this is the first method in the context of THMMs that controls false detection of a disturbed component.



We consider the lasso-penalised log-likelihood of model \eqref{mixtboats}, given by:
\begin{equation}
\label{mixtboatspenalide}
\begin{aligned}
\ell_p(\boldsymbol{\theta},{\boldsymbol{\beta}_0};\lambda)=\ell(\boldsymbol{\theta},{\boldsymbol{\beta}_0})- \lambda \lVert{\boldsymbol{\beta}_0}\lVert_1
\end{aligned},
\end{equation}
where $\lVert \cdot \rVert_1$ denotes the $\ell_1$-norm,
$\ell(\boldsymbol{\theta}, \boldsymbol{\beta}_0)
= \log \mathcal{L}(\boldsymbol{\theta}, \boldsymbol{\beta}_0)$,
and $\lambda \ge 0$ is a tuning parameter controlling the strength of the $\ell_1$ penalty. Essentially, for a large penalty strength, lasso regularisation shrinks the elements of $\boldsymbol{\beta}_0$ associated with unsupported disturbance effects towards zero, thereby preventing the inclusion of a disturbed regime when no disturbance is detectable. 

Observe that any $\boldsymbol{\beta}_0$ satisfying $
\text{(a)} 
 \quad \beta_0^i \leq  \frac{1}{\underset{t = 1,\ldots, T}{\max} u_t^i}  \ \forall \ i, \quad \text{ or}\quad \text{(b)}
 \quad  \frac{1}{\underset{t = 1,\ldots, T}{\min}u_t^i} \leq \beta_0^i , \ \forall \ i,
\quad
$ leads to a single-component model. These conditions specify that the mixture probability is constant over time whenever the threshold lies strictly outside the observed range of the covariate (above its maximum or below its minimum). The lasso penalty pushes elements satisfying condition (a) towards zero, such that $\nu_{\boldsymbol{0}}(\cdot)$ represents the baseline behaviour. Condition (b) is discouraged in practice, as the model is parametrised such that sufficiently low covariate values (e.g., low exposure) are associated with the baseline regime $B$. Such a constraint is reasonable in many applications, as there are typically conditions under which the process is known to be undisturbed. Consequently, the baseline component is always identifiable, and any single-component solution necessarily corresponds to $\nu_{\boldsymbol{0}}(\cdot)$.

In practice, selecting an appropriate penalty strength $\lambda$ in equation \eqref{mixtboatspenalide} is difficult. 
Current methods generally rely on grid searches across potential $\lambda$ values, using either cross-validation or information criteria for evaluation. These procedures can be computationally expensive, as each candidate value requires fitting one or several HMMs. While cross-validation is not straightforward to apply to time-series data due to temporal dependencies, dedicated strategies have been developed to account for this structure (\citealp{celeux_selecting_2008}; \citealp{roberts_crossvalidation_2017}). However, cross-validation methods further increase the computational cost of grid search procedures. Information criteria are computationally cheaper but lack theoretical justification in this setting because the null model (no disturbance) is unidentifiable and therefore has no unique MLE (see Section \ref{ple}). 
In Section {\ref{inference}}, we introduce a computationally efficient method to conduct inference to estimate both parameters and hyperparameters.

\subsubsection{qREML approach to select the penalty parameter}
\label{inference}
We propose a tractable and computationally efficient approach to select the penalty parameter of the lasso-penalised THMM, by treating the elements of $\boldsymbol{\beta}_0$ as random effects with a joint exponential distribution as their prior. This approach has been successfully applied to spline-based nonparametric HMMs, where the smoothing parameter is estimated via marginal maximum likelihood by integrating out Gaussian-distributed random effects (\citealp{michelot_hmmtmb_2023, koslik_efficient_2024}). We extend it here to select the lasso penalty parameter $\lambda$.

Treating $\boldsymbol{\beta}_0$ as a random effect with exponential prior $f_{\lambda}$, the marginal likelihood of the data $\boldsymbol{Y}$ as a function of $\boldsymbol{\theta}$ and $\lambda$ is
\begin{equation}
    \label{marginal}
    \mathcal{L}_p(\boldsymbol{\theta},\lambda)=\int f_{\boldsymbol{\theta}}(\boldsymbol{Y}| \boldsymbol{\beta}_0)f_{\lambda}(\boldsymbol{\beta}_0) \, d\boldsymbol{\beta}_0,
\end{equation}
where $f_{\lambda}$ is the prior distribution of $\boldsymbol{\beta}_0$ and  $f_{\boldsymbol{\theta}}(\boldsymbol{y}| \boldsymbol{\beta}_0)$ is the likelihood as a function of $\boldsymbol{\theta}$ and $\boldsymbol{\beta}_0$. 

Standard lasso penalisation corresponds to independent Laplace priors on $\boldsymbol{\beta}_0$ (\citealp{tibshirani_regression_1996}). In our case however, every element of $\boldsymbol{\beta}_0$ is assumed to be nonnegative, hence $f_{\lambda}(\boldsymbol{\beta}_0)$ is the joint distribution of independent exponential random variables with rate $\lambda$. Direct optimisation over both $\boldsymbol{\theta}$ and $\lambda$ via Laplace approximation of equation~\eqref{marginal} is computationally costly, so, following \cite{laird1982random} and \cite{koslik_efficient_2024}, we adopt a fully Bayesian framework with a non-informative multivariate normal prior on $\boldsymbol{\theta}$. Integrating out $\boldsymbol{\theta}$ and $\boldsymbol{\beta}_0$ and applying a Laplace approximation around the mode $(\boldsymbol{\hat{\theta}},\boldsymbol{\hat{\beta}}_0)$ yields the following marginal log-likelihood of $\lambda$ (the complete derivation is provided in the Supplementary Material):
\begin{equation}
\label{laplace}
\ell_p(\lambda)
= \ell(\boldsymbol{\hat{\theta}},\boldsymbol{\hat{\beta}}_0)
+ p_2 \log(\lambda)
- \lambda \lVert \boldsymbol{\hat{\beta}}_0 \rVert_1
- \frac12 \log\!\left( |\widehat{H}_{\lambda}| \right),
\end{equation}
where $\widehat{H}_{\lambda}$ is the negative Hessian matrix of
$
h_{\lambda}(\boldsymbol{\theta},\boldsymbol{\beta}_0)
= \ell(\boldsymbol{\theta},\boldsymbol{\beta}_0)
+ p_2 \log \lambda
- \lambda \lVert \boldsymbol{\beta}_0 \rVert_1
$
with respect to $(\boldsymbol{\theta},\boldsymbol{\beta}_0)$, evaluated at the
mode of $\ell_p(\cdot,\cdot;\lambda)$, and $|\cdot|$ denotes the determinant operator.

For the outer optimisation step, we follow \cite{koslik_efficient_2024} and differentiate equation~\eqref{laplace} with respect to $\lambda$ while treating $(\boldsymbol{\hat{\theta}},\boldsymbol{\hat{\beta}}_0)$ as fixed. The Hessian log-determinant term vanishes upon differentiation (see Supplementary Material), giving the following closed-form update
\begin{equation}
   \label{boom} 
    \lambda = \frac{p_2}{\sum_{i=1}^{p_2} {\hat{\beta}_{0}^i}} > 0.
\end{equation}
Since both sides of equation~\eqref{boom} depend on $\lambda$, we apply this update iteratively: at each step we compute the mode of the penalised log-likelihood via direct numerical optimisation (\citealp{zucchini_hidden_2017,koslik_efficient_2024}) for the current $\lambda$, then update $\lambda$ accordingly until convergence. All the derivation details on the qREML approach are provided in the Supplementary Material.

\subsection{Simulation}
We first demonstrate the proposed method's performance via a simulation study, verifying that it accurately recovers true disturbance thresholds when disturbance is present and shrinks them towards zero when it is not. We consider five scenarios involving one or two covariates: single-covariate settings with and without disturbance, and bivariate settings with both, one, or neither covariate linked to disturbance. Together, these scenarios assess the method's ability to accurately estimate $(\boldsymbol{\theta},\boldsymbol{\beta}_0)$ and control false detection rates across a wide range of conditions.

Data are generated from a three-state THMM as defined in equation \eqref{mixtboats}. We consider three covariate sequences: a real-valued sequence $\{u_{1,t}\}_{t=1}^{T}$, a binary sequence $\{u_{2,t}\}_{t=1}^{T}$, and a bivariate sequence $\{\boldsymbol{u}_{3,t}\}_{t=1}^{T}$, defined in terms of $\{u_{1,t}\}_{t=1}^{T}$ and $\{u_{2,t}\}_{t=1}^{T}$. The single-covariate scenarios use $\{u_{1,t}\}_{t=1}^{T}$, while the bivariate scenarios use $\{\boldsymbol{u}_{3,t}\}_{t=1}^{T}$. As is common in movement modelling, we use gamma state-dependent distributions to realistically mimic animal speed patterns (slow, moderate, and fast), with high persistence of $0.9$ (i.e., the diagonal elements of $\boldsymbol{\Gamma}^{(B)}$ are set to $0.9$), gamma means $\mu_0 = (1, 4, 10)$, and shape parameters $s_0 = (1.5, 10, 12)$ (\citealp{zucchini_hidden_2017,pohle_selecting_2017}). Under disturbance, the persistence for the last two states is reduced to $0.7$ to illustrate increased switching behaviour. This parametrisation could reflect the decrease in movement persistence observed by \cite{tervo_stuck_2023} and \cite{delporte_varying_2025} in narwhal exposed to ship noise and airgun pulses.

We simulate $\{u_{1,t}\}_{t=1}^{T}$ as a deterministic, smooth, periodic time series to reproduce the behaviour of environmental covariates such as temperature (see Supplementary Material for more details). The bivariate covariate $\{\boldsymbol{u}_{3,t}\}_{t=1}^{T}$ is designed to capture interactions between $\{u_{1,t}\}_{t=1}^T$ and $\{u_{2,t}\}_{t=1}^{T}$. This setup mirrors our case study, where we consider the interaction between the exposure to the closest vessel and the presence of land. 
Specifically, we define 
$$
\boldsymbol{u}_{3,t} = 
\begin{cases}
(u_{1,t}, 0), & \text{if } u_{2,t} = 1, \\
(0, u_{1,t}), & \text{if } u_{2,t} = 0,
\end{cases}
$$
such that the binary covariate (with each value set to the previous one with probability $p$) controls the disturbance threshold, assigning one threshold when $u_{2,t} = 1$ and another when $u_{2,t} = 0$.  

In Scenario 1, we explore two settings with covariate sequence $\{u_{1,t}\}_{t=1}^{T}$ and sample sizes
$T \in \{1{,}000, 3{,}000, 5{,}000, 10{,}000\}$. The sample sizes were chosen to reflect typical values reported in animal movement studies (\citealp{isojunno_individual_2017,shuert_putting_2025}). 
Scenario 1$.$a corresponds to data generated with an active threshold fixed at $21$ before data standardisation, resulting in disturbance frequencies of $0.62$, $0.35$, $0.50$, and $0.46$ for $T = 1{,}000$, $3{,}000$, $5{,}000$, and $10{,}000$, respectively. We additionally explored settings with less frequent disturbance, corresponding to the 5th percentile of the covariate distribution (results are provided in the Supplementary Material). Scenario 1$.$b simulates no disturbance effect and is therefore equivalent to a standard HMM (i.e., null model). In scenario 2, three configurations are explored with covariate sequence $\{\boldsymbol{u}_{3,t}\}_{t=1}^{T}$ and sample size $10{,}000$: scenario 2$.$a, where distinct thresholds are used for each covariate dimension ($21$ for dimension 1 and $30$ for dimension 2, corresponding to ($1.90$,$1.33$) after scaling); scenario 2$.$b, where only one of the two covariate dimensions is associated with a disturbance; and scenario 2$.$c corresponding to a standard HMM with no disturbance effect. 

To assess spurious detection of a disturbed regime under the null model (i.e., scenarios 1$.$b, 2$.$b, and 2$.$c), we define an empirical detection criterion based on the estimated mixture probability ${\nu}_{{\boldsymbol{\hat{\beta}}}_0}(t)$. Specifically, we consider a model fit to exhibit a spurious detection of disturbance if ${\nu}_{{\boldsymbol{\hat{\beta}}}_0}(t) > 0.001$ for any time point $t$.
This threshold reflects a conservative tolerance for non-zero transition probability to the disturbed regime under the null model and is used as a diagnostic criterion rather than as part of a formal hypothesis test. The resulting proportion of datasets exhibiting spurious detections provides an empirical false positive rate, which we compare to the frequency of detections obtained using the BLRT at the conventional $0.05$ significance level. Although formal post-model-selection inference is challenging (\citealp{zhang_post-model-selection_2022}), our goal here is simply to demonstrate the method’s reliability via simulation, using the false positive rate as a performance metric for false detection. 

We simulated $50$ datasets for each scenario and sample size. For each simulated dataset, we fit the lasso-penalised THMM and also perform a BLRT for comparison. To mitigate the risk of converging to local maxima, we initialise the optimisation from $50$ random starting values and select the fit with the highest likelihood. 
Simulations were run on the Cedar and Narval Compute Canada clusters, each with 15 CPUs and 8 GB of dedicated memory. Computational costs were estimated for a single model run. Since both models were run on 15 cores, the total runtime was divided by the number of initial values explored ($50$) and the number of cores to obtain the approximate cost for one run. The full reproducible simulation code is available on github (\url{https://github.com/Fanny-Dupont/THMM}). 



\subsection{Simulation results}
\label{simresult}
\begin{figure}[H]
\hspace{1.5cm}
    \includegraphics[width=13.5cm,height=5cm]{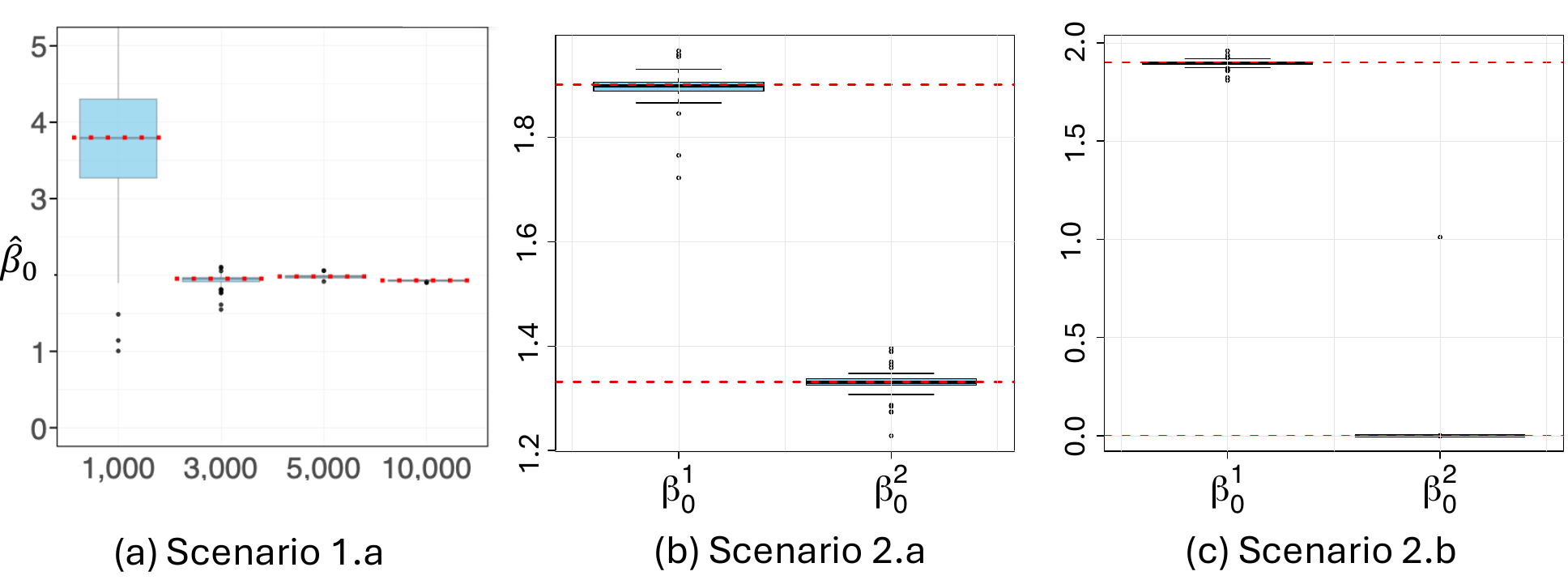}
    \caption{Estimates of $\boldsymbol{\beta}_0$ obtained using the lasso-penalised THMM. (a) Univariate setting in the presence of disturbances (scenario 1$.$a) across different sample sizes; the red dotted lines correspond to the true value of $\beta_0$ for each sample size. (b)--(c) Bivariate setting, $\boldsymbol{\beta}_0 = (\beta_0^1,\beta_0^2)$, with sample size $10{,}000$: (b) both covariates have different disturbance thresholds, $\boldsymbol{\beta}_0 = (1.90,1.33)$ (scenario 2$.$a); (c) only one covariate has an active threshold, $\boldsymbol{\beta}_0 = (1.90,0)$ (scenario 2$.$b). The red dotted lines correspond to the true value of each element of $\boldsymbol{\beta}_0$.}

    \label{result1}
\end{figure}

Our method provided accurate estimates of $\boldsymbol{\beta}_0$ across all scenarios while being substantially faster than the BLRT (see Table \ref{tabtab}). For sample sizes larger than $3{,}000$, the lasso-penalised THMM achieves excellent control of false positive rate (below $0.02$), and always identifies disturbances when they occur (Figure \ref{result1}). The mean of the $\boldsymbol{\beta}_0$ estimates across the simulated datasets matches the simulated true values (Figure \ref{result1}). Additionally, the estimated state-dependent parameters exhibit low absolute bias and variance, with estimates closely aligning with the simulated values (see Table \ref{bias}). The largest bias observed corresponds to 2.7\% of the true parameter value.

While the simulated disturbance threshold is fixed at $21$ in Scenario~1.a, the true value of $\boldsymbol{\beta}_0$ varies with sample size due to covariate standardisation. Our method yields accurate estimates of $\boldsymbol{\beta}_0$, with bias and variance decreasing as the sample size increases (Figure~\ref{result1}a). For sample sizes $T \in \{1{,}000, 3{,}000, 5{,}000, 10{,}000\}$, the corresponding bias values are
$0.54$, $-0.02$, $-0.004$, and $-0.003$, indicating that the model reliably detects a true disturbance effect and that bias is negligible for $T \ge 3{,}000$. Estimation precision also improves with sample size, as shown by the declining standard deviations of $3.0,0.10,0.026$, and $0.008$. In the presence of disturbance (i.e., scenario 1$.$a), the distribution of $\hat{\lambda}$ is highly concentrated around its mean (Figure \ref{fig:lambda1}).

Under the null model, corresponding to scenario 1$.$b, most $\hat{\lambda}$ values are very large, effectively shrinking ${\hat{\beta}}_0$ towards zero. However, occasional outliers ($\hat{\lambda} < 1$) can occur. For sample sizes $T \geq 3{,}000$, these outliers lead to ${\hat{\beta}}_0$ values near the standardised maximum ($\max_t \text{ }{u}_t = 1$; Figs \ref{result1}c, \ref{resultsold1}b and \ref{resultsold2}b-c) and the estimated thresholds (${1}/{{\hat{\beta}}_0}$) lie entirely above the range of observed covariate values, effectively indicating no disturbance (more details in Section \ref{modeldescription}) as shown by the estimated false positive rate of $0$ (Table \ref{tab:typeIerror1}). These outliers likely stem from the Laplace approximation’s reliance on posterior normality. While this assumption is theoretically supported by the asymptotic properties of the MLE for large sample sizes (\citealp{jensen_asymptotic_2011}), the approximation becomes less accurate for moderate sample sizes (e.g., $T = 1{,}000$), failing to enforce the lasso’s expected behaviour of shrinking spurious $\boldsymbol{\hat{\beta}}_0$ towards zero under the null. As a result, the proportion of outliers grows with decreasing sample size and the rate of false positives increases. However, this bias does not appear to substantially affect other parameter estimates. Indeed, biases for $\boldsymbol{\theta}$ under the null and disturbed scenarios are very similar, suggesting that any bias in the Laplace approximation under the null model primarily impacts the estimate of $\boldsymbol{\beta}_0$ and does not propagate to the remaining parameters. This pattern likely arises because the elements of $\boldsymbol{\theta}$ are unpenalised and therefore not directly influenced by the bias introduced by the Laplace approximation.

\begin{table*}[ht]
\caption{False positive error rates and computational costs (first and third quartiles in minutes) for BLRT and lasso-penalised THMMs, scenario 1.b (null model).}
\label{tab:typeIerror1}
\small
\setlength{\tabcolsep}{4pt}
\begin{tabular}{lcccccccc}
\toprule
 & \multicolumn{4}{c}{BLRT} & \multicolumn{4}{c}{Lasso} \\
\cmidrule(lr){2-5} \cmidrule(lr){6-9}
$T$& $1{,}000$ & $3{,}000$ & $5{,}000$ & $10{,}000$ & $1{,}000$ & $3{,}000$ & $5{,}000$ & $10{,}000$ \\
\midrule
False positive rate & 0.10 & 0.13 & 0.07 & 0.03 & 0.20 & 0.02 & 0.00 & 0.00 \\
Computational cost & 3.2--3.3 & 6.4--7.3 & 10.4--11.1 & 20.7--22.9 & 2.6--3.1 & 3.2--4.8 & 6.2--9.4 & 10.2--12.7 \\
\bottomrule
\end{tabular}
\end{table*}




In bivariate settings, our method provides accurate estimation of $\boldsymbol{\beta}_0$ for both covariates (Figure \ref{result1}b-c), with false positive rates comparable to or lower than the BLRT's (see Supplementary Material). 
In Scenario 2$.$b, our method exhibits a low false positive rate of $0.02$. In contrast, the BLRT performs poorly when identifying the covariate responsible for the disturbance: it selects the incorrect covariate in $48\%$ of cases and correctly identifies the true covariate only $70\%$ of the time. Our approach substantially outperforms BLRT in reliably attributing the disturbance effect, consistently identifying the correct disturbance covariate across simulations (Figure~\ref{result1}b-c). We believe our method performs well in scenario 2$.$b because it can use the information from the disturbed covariate to identify disturbance patterns and clearly distinguish between regimes. 
Under Scenario~2$.$c, both methods exhibit higher false positive rates than in the univariate case, likely due to reduced effective sample size per covariate and increased model complexity (see Table \ref{tabtab}). In this setting, our method yields false positive rates of $0.06$ and $0.15$, compared to $0.12$ for BLRT.

\section{Application to narwhal movement data}
\subsection{Investigating the effect of vessel presence}

When applying our method to narwhal movement data, our objectives are to estimate the distance at which narwhal respond to vessels and assess whether disturbance thresholds differ depending on whether land lies between the whale and the vessel.

We fit a THMM in which the covariate sequence $\{\boldsymbol{u}_t\}_{t=1}^T$ is the interaction between \textit{presence of land} and \textit{exposure}, where exposure is defined as the inverse distance (in km$^{-1}$) between the whale and the nearest vessel. Accordingly, as in scenario~2$.$b, we define
$$
\boldsymbol{u}_{t} =
\begin{cases}
(\text{exposure}_{t}, 0), & \text{if land lies between the whale and the vessel}, \\
(0, \text{exposure}_{t}), & \text{if no land lies between them}.
\end{cases}
$$
We also include distance to shore as a covariate in the TPMs for both regimes, since previous research has identified it as a key factor influencing narwhal behaviour (\citealp{heide-jorgensen_behavioral_2021,dupont2025improved, hornby2025behavioural}). To maintain model parsimony, we constrained the effect of distance to shore to be equal across both regimes.

Selecting the number of states in HMMs is challenging (\citealp{pohle_selecting_2017,dupont2025improved}) but here, we follow \cite{ngo_understanding_2019} and \cite{shuert_assessing_2021} and use three behavioural states, as this number has been shown to adequately capture the range of behaviours exhibited by this species in similar studies. We use independent gamma distributions for maximum depth and step length, and a von Mises distribution for turning angles as is typical in animal movement studies (\citealp{mcclintock_momentuhmm_2018,shuert_putting_2025,hornby2025behavioural}). While independent parametric state-dependent distributions are standard in animal movement analyses, our framework can support more complex distributions such as cylindrical distributions or nonparametric spline-based alternatives (\citealp{lagona2015hidden, langrock_nonparametric_2015,abe2017tractable}). Narwhal positions more than $77$ km from the nearest vessel were assigned to the baseline behaviour. This cutoff corresponds to the $60$th percentile of the distribution of distances to the nearest vessel, which is reasonable given that prior studies have documented disturbance responses at distances of up to $40$ km (\citealp{heide-jorgensen_behavioral_2021}). We explored a range of candidate baseline cutoffs and found that the negative log-likelihood favoured 77 km over the other values tested (see Supplementary Material for details). We used $100$ random initial values to reduce the risk of convergence to a local minimum (\citealp{zucchini_hidden_2017}; \citealp{mcclintock_worth_2021}). Bootstrap confidence intervals for the threshold estimates were obtained by resampling the 231 tracks with replacement and refitting the full model on each bootstrap sample. This approach allows us to assess the variability in threshold estimation as well as the frequency with which disturbance is detected across resamples. A tutorial with complete, reproducible code for the case study, along with the data used to generate the results, is available on GitHub (\url{https://github.com/Fanny-Dupont/THMM}).

\subsection{Results}
Our results suggest that the presence of vessels affects narwhal behaviour. Narwhal seem to react to vessels up to $3.4$ km away (95\% bootstrap CI: $0.33$--$11.66$ km, with disturbance detected in $95\%$ of bootstrap resamples) by decreasing movement persistence and spending more time in deep water. No behavioural changes were detected when land separates the whale from the nearest vessel. The lack of behavioural change in the presence of land likely arises because islands and peninsulas block vessel noise, which limits acoustic exposure.

\begin{figure}[H]
    \centering
    \hspace{-1.5cm}
    \includegraphics[width=16cm,height=5.5cm]{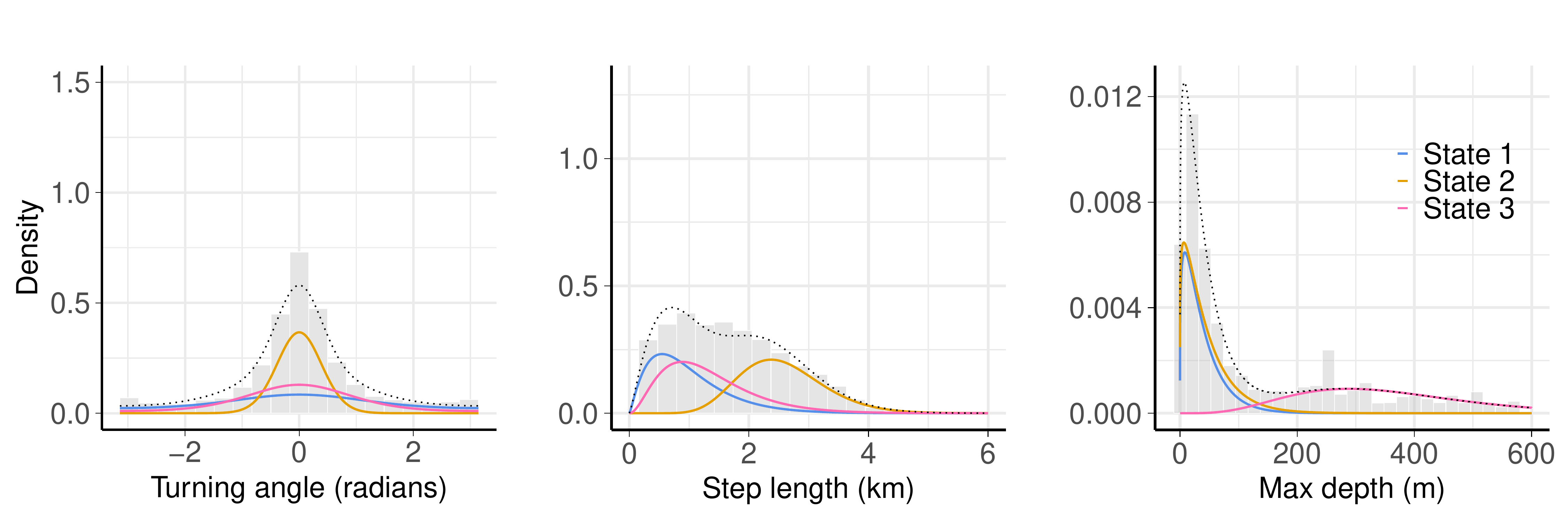}
    \caption{Estimated distributions from the three-state THMM. Each colour corresponds to a different state.}
    \label{fig:top}
\end{figure}

\begin{figure}[H]
    \centering
    \includegraphics[width=9cm,height=5.5cm]{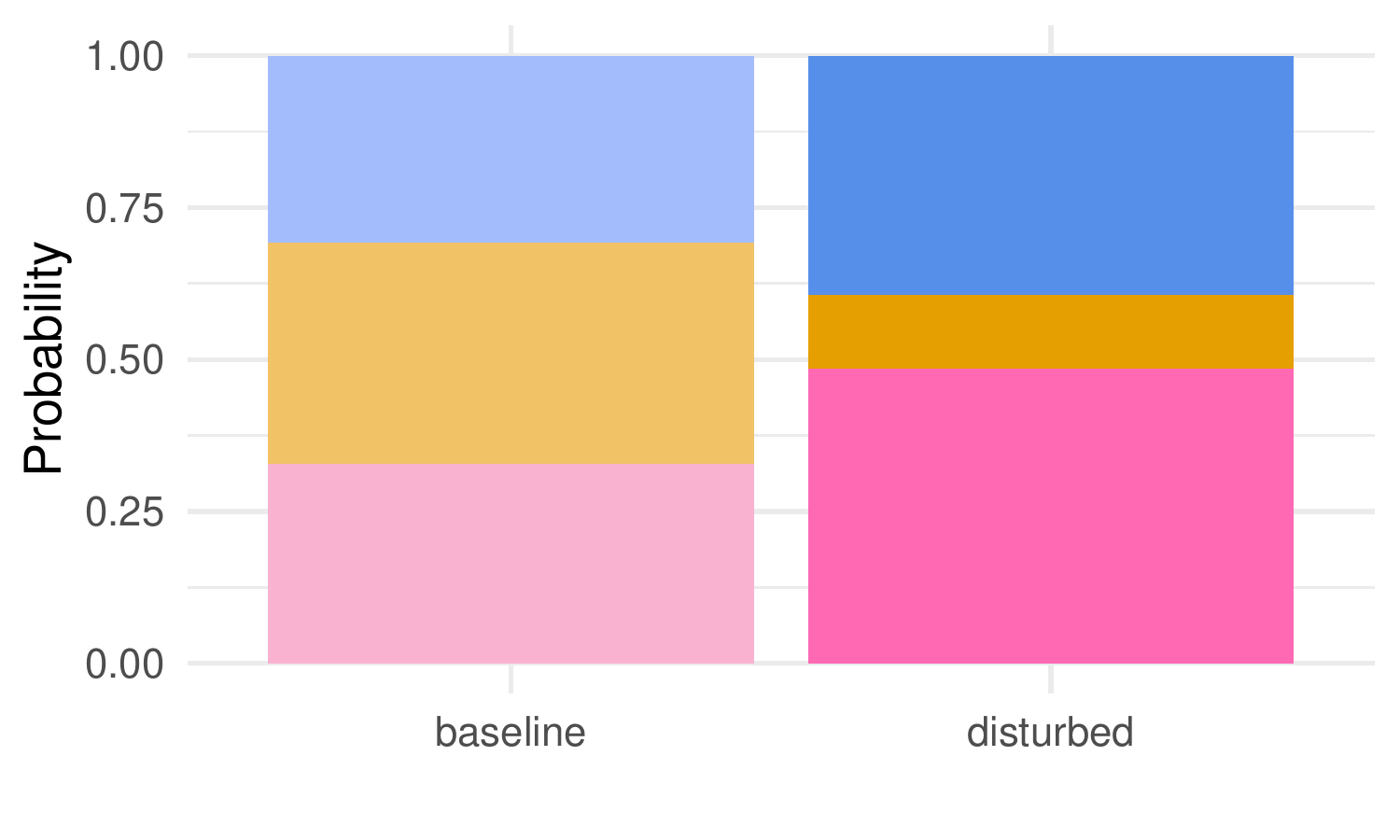}
    \caption{Percentage of time in each state for baseline and disturbed components.}
    \label{fig:bottom}
\end{figure}

The three estimated states correspond to distinct behaviours (Figure \ref{fig:top}). State 1 corresponds to slow (average step-length of $1$km), undirected, shallow movement, state 2 is associated with fast (average step length of $2.6$km), directed, shallow movement, and state 3 with deep, undirected, and slow movement (average step length of $1.3$ km; see Supplementary Material for all state-dependent parameter estimates). Using the Viterbi algorithm (\citealp{forney_viterbi_1973}), we derived the time allocation to each state, revealing that narwhal spend approximately two-thirds of their time in surface-associated behaviours (states 1–2; Figure \ref{fig:bottom}) in the absence of disturbance. This pattern is consistent with findings by \cite{watt_differences_2015}, which report predominant shallow-water activity.

The model indicates that narwhal respond to vessel presence at distances of up to approximately $3.4$ km, consistent with findings from \cite{golder_maryriver_2020}, who reported significant behavioural changes occurring within $1$-$4$ km of vessels. Within $3.4$ km from the nearest vessel, our results show changes in the state transition probabilities, with magnitude depending on the animal’s distance to shore. Specifically, behavioural responses are stronger farther from shore. At a distance of $5.43$ km from shore (the mean distance to shore during disturbance), the probability of transitioning from slow, shallow movement (state 1) to deep diving (state 3) increases substantially (from $0.13$ to $0.33$), while persistence in state 1 decreases from $0.75$ to $0.64$. The increased diving observed in response to vessel noise (Figure \ref{fig:bottom}) resembles the escape behaviour documented during killer whale encounters, in which narwhal increase dive frequency and exhibit prolonged submergence (\citealp{williams2011extreme, breed_sustained_2017}). In contrast, at shallow depths close to shore (first quartile of distance to shore, $0.65$ km), changes in those states are more limited: persistence in state 1 decreases slightly from $0.79$ to $0.77$, while persistence in the deep-diving state remains unchanged. These weaker behavioural responses nearshore likely reflect spatial and bathymetric constraints that limit deep diving. 
Regardless of distance to shore, persistence in state~2 (directed and fast movement) declines under disturbance, decreasing from $0.82$ in the baseline regime to $0.54$ in the disturbed regime (evaluated at the first and third quartiles of distance to shore). This result is consistent with \cite{delporte_varying_2025}, who found that narwhal exhibit decreased movement persistence (i.e., decrease in directed movement) when exposed to vessels. 

\section{Discussion}
Threshold HMMs extend traditional HMMs to simultaneously model baseline behaviour and detect the point along a continuous covariate (e.g., exposure to a vessel) where behaviour shifts noticeably from baseline patterns. We introduced a novel, computationally efficient lasso-penalised THMM to estimate disturbance thresholds in narwhal movement data while controlling for false detection. Our method, based on a fast and efficient qREML approach for penalty selection, offers both computational speed and reliable results. It accurately estimates movement parameters, differentiates baseline from disturbed behaviours, and reliably captures disturbance thresholds. A key advantage of lasso-penalised THMMs is their ability not only to estimate disturbance thresholds but also to shrink spurious disturbance effects towards zero when there is no measurable disturbance, thereby providing rigorous evidence of behavioural changes and addressing a gap in the existing literature.
We believe that our method represents a crucial step towards establishing new, rapid model selection techniques for THMMs, eliminating the need for computationally expensive grid searches to determine penalty strength.

Our analysis of narwhal movement data reveals that our method can successfully identify complex behavioural responses to disturbance.
We presented the first model-based estimates of disturbance thresholds from anthropogenic activities. 
Our findings contribute to the increasing evidence that human activity affects the behaviour of Arctic marine mammals (\citealp{tervo_narwhals_2021}; \citealp{williams_physiological_2022}; \citealp{tervo_stuck_2023}; \citealp{martin_exposure_2023}; \citealp{delporte_varying_2025}; \citealp{ewing2025narwhal}). Our results support previous findings that narwhal respond to vessels  (\citealp{golder_maryriver_2020,tervo_narwhals_2021,delporte_varying_2025,ewing2025narwhal}), and align with the findings by \cite{golder_maryriver_2020}, which showed that behavioural reactions occurred within $4$ km of vessels. Our analysis suggests that narwhal exposed to vessels spend more time at depth and exhibit reduced movement persistence. The divergence from \cite{tervo_narwhals_2021} likely reflects methodological differences (i.e., controlled sound exposure experiment using airgun pulses). Our findings share similarities with \cite{delporte_varying_2025} who reported a decrease in movement persistence in narwhal under disturbance and with the ``escape dive" behaviour observed by \cite{williams2017paradoxical}, potentially indicating related avoidance strategies. Such avoidance strategies have also been observed in contexts of predator evasion, with narwhal demonstrating prolonged submergence to escape killer whales (\citealp{williams2011extreme}) and performing deep dives 
more frequently (\citealp{breed_sustained_2017}). This behaviour is likely costly and may impact their energy budgets (\citealp{williams2017paradoxical}). Consequently, narwhal may need to increase foraging efforts to restore energy balance. However, deep diving may also correspond to foraging behaviour (\citealp{shuert_putting_2025}), and high frequency acceleration data would help investigate behaviour at depth. Interpreting behavioural responses is challenging, given the complexity of narwhal behaviour and the limited data available. This difficulty highlights the need for additional research, to validate and build upon our findings. In principle, one could further investigate whether the detected disturbance differs among vessel types. However, we could not explore this variation because the estimated threshold resulted in too few observations in the disturbed regime to support additional covariates.


Given the extensive use of both likelihood ratios and bootstrap techniques in statistical analysis, the BLRT stands out as a particularly attractive alternative to our method (\citealp{mclachlan_bootstrapping_1987}; \citealp{dziak_effect_2014}). However, the method’s computational cost is a notable limitation that may become prohibitive as ecological time-series datasets grow in size (\citealp{patterson_statistical_2017}). Another difficulty arises when testing multiple thresholds since defining an appropriate null model is ambiguous. For example, in our narwhal analysis, we would have to choose whether the null hypothesis for the BLRT assumed a complete absence of effects, or the absence of disturbance effects only in the presence of land. Such distinctions are non-trivial, as each null hypothesis requires fitting a separate model, and no single test accommodates multiple null specifications simultaneously. Testing many null hypotheses becomes impractical as model complexity increases. For example, including additional factors, such as vessel category, would lead to a large number of null models that must be evaluated separately. 
The lack of a single, well-defined null hypothesis complicates the use of the current BLRT approach, as it requires either a carefully constructed null hypothesis or multiple tests (using different null models), which would substantially increase computational costs and introduce challenges related to multiple testing and type I error rate inflation.

While the lasso approach provides effective variable selection, its performance depends on the accuracy of the Laplace approximation (via the chosen penalty strength), and the extent of this dependence is not yet fully understood. Future work could focus on incorporating post-selection inference techniques to construct confidence intervals and uncertainty measures (\citealp{zhang_post-model-selection_2022}). Such extensions would enable more comprehensive statistical inference while maintaining the computational advantages of the method. In our analysis, we use bootstrap confidence intervals for the threshold estimates, which provide a practical measure of uncertainty. However, their validity is complicated by the temporal dependence structure of the data. In particular, the choice of resampling unit is non-trivial for time-series data with few independent trajectories, and resampling split tracks does not fully account for within-track dependence. Overall, the approach we propose offers an efficient and reliable method to assess whether the detected disturbance in THMMs is meaningful. Importantly, our method generalises beyond using THMMs to detect response to disturbance since THMMs can be used to quantify any reaction to a stimulus. The THMM framework is highly adaptable, capable of accommodating various types of threshold analysis in time-series data beyond the scope of animal movement.

In telemetry studies, THMMs can be used to estimate fundamental thresholds that have been difficult to characterise, such as the distance at which elephants can detect water (\citealp{wood_african_2022}) or the distance at which bowhead whales react to killer whales (\citealp{matthews2020killer}). However, the use of THMMs and standard HMMs in ecology is not restricted to movement data (\citealp{zucchini_hidden_2017}). Threshold HMMs could be applied to estimate the temperature above which coral bleaching becomes significantly more severe and recovery is unlikely (\citealp{kayanne2017validation}). Our framework naturally extends to other latent variable models with threshold effects, including state-space models or THMMs with random effects (\citealp{altman_mixed_2007}; \citealp{augermethe_guide_2021}; \citealp{mcclintock_worth_2021}). This flexibility suggests promising directions for future methodological developments and applications.

\section*{Acknowledgments}
We thank Jan-Ole Koslik, Dr. Matías Salibián-Barrera, and Dr. Nancy E. Heckman for their insights and feedback. We thank the community of Mittimatalik (Pond Inlet) for its support in tagging operations and the devoted people who led operations in the field.

\section*{Funding}
We would like to thank the Natural Sciences and Engineering Research Council of Canada (NSERC), Canada Research Chairs program, BC Knowledge Development Fund, Canada Foundation for Innovation's John R. Evans Leaders Fund, Canadian Statistical Sciences Institute (CANSSI), Fisheries and Oceans Canada (DFO), and the  Arctic Section of the Society of Naval Architects and Marine Engineers for their support. Fieldwork was supported by the Polar Continental Shelf Program, DFO, the Nunavut Wildlife Management Board, the Nunavut Implementation Fund, World Wildlife Fund Canada. This research was enabled by support provided by Compute Canada (www.alliancecan.ca).


\bibliographystyle{abbrvnat}
\bibliography{bibli} 

\newpage 
\appendix

\renewcommand\thefigure{A\arabic{figure}} 
\renewcommand\thetable{A\arabic{table}} 
\setcounter{figure}{0} 
\setcounter{table}{0} 
\section{Supporting Information}
\subsection{Details of the qREML derivation for the lasso penalty parameter}
\label{supp:qreml}

\subsubsection{Marginal log-likelihood via Laplace approximation}
The main idea is to treat the vector of coefficients $\boldsymbol{\beta}_0$ as a random effect and consider
the marginal likelihood of the data $\boldsymbol{y}$, as a function of $\boldsymbol{\theta}$ and $\lambda$ as in equation \eqref{marginal}. The penalty term in equation \eqref{mixtboatspenalide} can be interpreted as the logarithm of the distribution for $\boldsymbol{\beta}_0$ (\citealp{koslik_efficient_2024,michelot_hmmtmb_2023}). Standard lasso penalisation is equivalent to using independent Laplace priors on the elements of $\boldsymbol{\beta}_0$ in a Bayesian formulation (\citealp{tibshirani_regression_1996}). In our case however, every element of $\boldsymbol{\beta}_0$ is assumed to be nonnegative, hence $f_{\lambda}(\boldsymbol{\beta}_0)$ is the joint distribution of independent exponential random variables with rate $\lambda$. While the integral in equation \eqref{marginal} can be approximated via Laplace approximation (\citealp{erkanli1994laplace, vandervaart}), the nested optimisation (over both $\boldsymbol{\theta}$ and $\lambda$) is computationally costly. Following \cite{laird1982random} and \cite{koslik_efficient_2024}, we extend our approach to a fully Bayesian framework by assigning prior distributions to all parameters. We use a multivariate normal prior with a sufficiently large variance on $\boldsymbol{\theta}$. From a Bayesian perspective, this corresponds to a non-informative (flat) prior and therefore imposes no penalty on $\boldsymbol{\theta}$.
The variance is chosen to be large enough such that the prior density is effectively constant over the region of parameter space supported by the likelihood. Consequently, its precise value is not of interest and is not estimated. Empirical simulation results (Section~\ref{simresult} and Supplementary Material) confirm that omitting this
normal prior from the posterior induces negligible bias when the prior precision is sufficiently low. Consequently, the resulting estimates of 
$\boldsymbol{\theta}$ are close to the maximum likelihood estimates and should exhibit minimal bias.

Starting from the marginal likelihood in equation~\eqref{marginal} of the main text, we integrate out both $\boldsymbol{\theta}$ and $\boldsymbol{\beta}_0$:
\begin{align}
\label{prelaplace}
    \mathcal{L}_p(\lambda)    &= \int f_{\boldsymbol{\theta}}(\boldsymbol{Y}| \boldsymbol{\beta}_0) f_{\lambda}(\boldsymbol{\beta}_0) \, d\boldsymbol{\beta}_0 \, d\boldsymbol{\theta} \notag \\
     &= \int \mathcal{L}(\boldsymbol{\theta},\boldsymbol{\beta}_0) f_{\lambda}(\boldsymbol{\beta}_0) \, d\boldsymbol{\beta}_0 \, d\boldsymbol{\theta} \notag \\
    &= \int e^{\ell(\boldsymbol{\theta},\boldsymbol{\beta}_0)} e^{\log(f_{\lambda}(\boldsymbol{\beta}_0))} \, d\boldsymbol{\beta}_0 \, d\boldsymbol{\theta} \notag \\
    &= \int e^{\ell(\boldsymbol{\theta},\boldsymbol{\beta}_0)} e^{{p_2} \log \lambda} e^{-\lambda \sum_{i=1}^{p_2} \beta_{0}^i} \, d\boldsymbol{\beta}_0 \, d\boldsymbol{\theta}.
\end{align}
The equivalence $f_{\boldsymbol{\theta}}(\boldsymbol{Y}|\boldsymbol{\beta}_0) \equiv \mathcal{L}(\boldsymbol{\theta}, \boldsymbol{\beta}_0)$ reflects the Bayesian motivation for this derivation, while estimation itself follows the frequentist paradigm through maximum likelihood.

The integral in equation~\eqref{prelaplace} is intractable, so we apply a Laplace approximation around the mode $(\boldsymbol{\hat{\theta}}, \boldsymbol{\hat{\beta}}_0)$, justified for a fixed $\lambda$ by the Bernstein--von Mises theorem (\citealp{vandervaart}). Intuitively, in equation \eqref{prelaplace}, 
$e^{\ell(\boldsymbol{\theta}, \boldsymbol{\beta}_0)}$ is the likelihood of a nonhomogeneous HMM, and, under standard regularity conditions, the corresponding maximum likelihood estimator is asymptotically normal (see \cite{jensen_asymptotic_2011} for more details). For a fixed $\lambda$, the remaining term, corresponding to the prior on $\boldsymbol{\beta}_0$, becomes negligible relative to the likelihood as the sample size grows, and the posterior concentrates increasingly around the mode.

Dropping additive constants, this leads to the following approximation of the marginal log-likelihood of $\lambda$:

\begin{equation}
\label{laplaceSupp}
\ell_p(\lambda)
= \ell(\boldsymbol{\hat{\theta}},\boldsymbol{\hat{\beta}}_0)
+ p_2 \log(\lambda)
- \lambda \lVert \boldsymbol{\hat{\beta}}_0 \rVert_1
- \frac12 \log\!\left( |\widehat{H}_{\lambda}| \right),
\end{equation}
where $\widehat{H}_{\lambda}$ is the negative Hessian matrix of
$
h_{\lambda}(\boldsymbol{\theta},\boldsymbol{\beta}_0)
= \ell(\boldsymbol{\theta},\boldsymbol{\beta}_0)
+ p_2 \log \lambda
- \lambda \lVert \boldsymbol{\beta}_0 \rVert_1
$
with respect to $(\boldsymbol{\theta},\boldsymbol{\beta}_0)$, evaluated at the
mode, and $|\cdot|$ denotes the determinant operator.  

For the outer optimisation step, the objective is to maximize the marginal log-likelihood $ \ell_p(\lambda) $ with respect to $ \lambda $. Following \cite{koslik_efficient_2024}, we use a qREML approach by computing partial derivatives of equation \eqref{laplace} while treating $(\boldsymbol{\hat{\theta}},\boldsymbol{\hat{\beta}}_0)$ as fixed quantities, despite their dependence on $\lambda$. This approximation to the full REML solution yields the following partial derivative:

    \begin{equation}
        \frac{\partial}{\partial \lambda} \ell_p(\lambda) = \frac{p_2}{\lambda} - ||\boldsymbol{\hat{\beta}}_0||_1 - \frac{1}{2} \operatorname{tr}\left( \widehat{H}_\lambda^{-1} \frac{d}{d\lambda} \widehat{H}_\lambda \right).
    \end{equation}

The third term arises from differentiating the log-determinant in equation~\eqref{laplace}. By Jacobi's formula (\citealp[Chapter~8]{magnus2019matrix}), $\frac{d}{dx}\log |A(x)| = \operatorname{tr}\!\left\{ A(x)^{-1}\frac{d}{dx}A(x)\right\}$, which holds for symmetric matrices $A(x)$ for all $x$. In our setting this term simplifies to zero, because the $\ell_1$ penalty contributes no second derivative with respect to $\boldsymbol{\beta}_0$: after taking second derivatives to form the Hessian, the resulting matrix does not depend on $\lambda$, and its derivative with respect to $\lambda$ therefore vanishes. Setting the remaining derivative to zero gives the closed-form update in equation~\eqref{boom} of the main text.

\subsubsection{Hessian derivation}\label{hessderiv}

In this section, we derive the Hessian matrix used in equation (\ref{laplace}) and explain the simplification of the trace term appearing in the marginal likelihood approximation. Recall that $ \widehat{H}_{\lambda} $ denotes the negative Hessian of the function 
$
h_{\lambda}(\boldsymbol{\theta}, \boldsymbol{\beta}_0) = \ell(\boldsymbol{\theta}, \boldsymbol{\beta}_0) + p_2 \log \lambda - \lambda \|\boldsymbol{\beta}_0\|_1,
$
with respect to $(\boldsymbol{\theta}, \boldsymbol{\beta}_0) = (\theta^1,\ldots, \theta^{p_1}, {\beta}_0^1,\ldots,  {\beta}_0^{p_2}) \in \mathbb{R}_{}^{p_1}\times\mathbb{R}_{+}^{*,p_2}$, evaluated at the mode. Consequently, we obtain the following second-order derivatives:

$$
\left\{
\begin{array}{ll}
\left.\dfrac{\partial^2}{\partial \theta^i \partial \theta^j}
h_{\lambda}(\boldsymbol{\theta}, \boldsymbol{\beta}_0)
\right|_{(\boldsymbol{\hat{\theta}}, \boldsymbol{\hat{\beta}}_0)}
=
\left.\dfrac{\partial^2}{\partial \theta^i \partial \theta^j}
\ell(\boldsymbol{\theta}, \boldsymbol{\beta}_0)
\right|_{(\boldsymbol{\hat{\theta}}, \boldsymbol{\hat{\beta}}_0)},
\\[1cm]
\left.\dfrac{\partial^2}{\partial \beta_{0}^i \partial \theta^j}
h_{\lambda}(\boldsymbol{\theta}, \boldsymbol{\beta}_0)
\right|_{(\boldsymbol{\hat{\theta}}, \boldsymbol{\hat{\beta}}_0)}
=
\left.\dfrac{\partial^2}{\partial \beta_{0}^i \partial \theta^j}
\ell(\boldsymbol{\theta}, \boldsymbol{\beta}_0)
\right|_{(\boldsymbol{\hat{\theta}}, \boldsymbol{\hat{\beta}}_0)},
\\[1cm]
\left.\dfrac{\partial^2}{\partial \beta_{0}^i \partial \beta_{0}^j}
h_{\lambda}(\boldsymbol{\theta}, \boldsymbol{\beta}_0)
\right|_{(\boldsymbol{\hat{\theta}}, \boldsymbol{\hat{\beta}}_0)}
=
\left.\dfrac{\partial^2}{\partial \beta_{0}^i \partial \beta_{0}^j}
\ell(\boldsymbol{\theta}, \boldsymbol{\beta}_0)
\right|_{(\boldsymbol{\hat{\theta}}, \boldsymbol{\hat{\beta}}_0)},
\\[1cm]
\left.\dfrac{\partial^2}{\partial (\beta_{0}^i)^2}
h_{\lambda}(\boldsymbol{\theta}, \boldsymbol{\beta}_0)
\right|_{(\boldsymbol{\hat{\theta}}, \boldsymbol{\hat{\beta}}_0)}
=
\left.\dfrac{\partial^2}{\partial (\beta_{0}^i)^2}
\ell(\boldsymbol{\theta}, \boldsymbol{\beta}_0)
\right|_{(\boldsymbol{\hat{\theta}}, \boldsymbol{\hat{\beta}}_0)}.
\end{array}
\right.
$$

Since $\widehat{H}_{\lambda}$ does not depend on $\lambda$, its derivative with respect to $\lambda$ is the zero matrix.

\subsection{Numerical implementation} 
Likelihood maximization is carried out via direct numerical optimisation using gradient-based algorithms. We prefer this approach over the Baum–Welch algorithm and gradient-free methods such as Nelder–Mead, which are typically slower for complex models (\citealp{lagarias1998convergence, zucchini_hidden_2017, RCoreTeam}). Moreover, gradient-based optimisation can accommodate complex model structures and benefits from automatic differentiation, as implemented in modern \texttt{R} packages such as \texttt{RTMB} and \texttt{LaMa} (\citealp{kristensen_tmb_2016,lama}).

Recall that the step function $\nu_{\beta_0}(\cdot)$ is discontinuous and therefore not suitable to gradient-based optimisation. We thus approximate it by a smooth, two-parameter logistic function defined in equation \eqref{twoparamlogistic},
where the parameter $b$ controls the sharpness of the approximation, with larger values yielding a closer approximation to a step function. Note that for multivariate $\boldsymbol{\beta}_0$, we have
\begin{equation}
\label{twoparamlogisticbis}
\nu_{\beta_0}(u_t)
\approx \left[1 + \exp^{-b\left(\boldsymbol{\beta}_0^{\top}
\boldsymbol{u}_t - 1\right)}\right]^{-1},
\end{equation}
and $b>0$ is a single global smoothing parameter controlling the sharpness of
the approximation. A key challenge when fitting the THMM with the two-parameter logistic function is the pathological behaviour of the likelihood gradient, and consequently the gradient of the penalised log-likelihood defined in
equation \eqref{mixtboatspenalide}, for large $b$. Specifically, the gradient vanishes when $\boldsymbol{\beta}_0$ is far from the maximizer of the penalised log-likelihood, but explodes near the maximum due to the sharp transition introduced by the smoothed step function. Consequently, inference is unstable and heavily dependent on initial values, as observed by \cite{fong_chngpt_2017}. To overcome this difficulty, we implement a progressive sharpness initialization strategy. We first fit the null model to obtain an estimate of the parameters of the state-dependent distributions. Keeping these estimates fixed, we then maximize the likelihood of the unpenalised THMM over $\boldsymbol{\beta}_0$ and the parameters of the hidden process, while gradually increasing the sharpness parameter $b$. This procedure gradually zooms in on the optimal region without numerical instability. Finally, we fit the penalised model using the target sharpness parameter ($b > 500$). This value is large enough to accurately approximate the step function. The optimisation is initialized at  $\boldsymbol{\hat{\beta}}_0$, $\hat{\boldsymbol{\alpha}}^{(B)},\hat{\boldsymbol{\alpha}}^{(D)},\boldsymbol{\hat{\delta}}^{(D)}$ and $\boldsymbol{\hat{\delta}}^{(B)}$ with the estimate obtained from the progressive procedure. We also stabilize inference through two constraints. First, as outlined previously, we initialize the model such that the component associated with $\nu_{\boldsymbol{0}}(\cdot)$ is already assigned a portion of the data. Second, we require the following constraint for the algorithm $|\Gamma^{(B)}_{ii} - \Gamma^{(D)}_{ii}| \geq \epsilon>0$ for at least one state $i \leq N$, guaranteeing measurable regime differentiation. In our implementation, we set $ \epsilon = 0.15 $, though this value can be modified (see code for details). 
To improve computational efficiency and numerical accuracy, we use the packages \texttt{RTMB} (\citealp{kristensen_tmb_2016}) and \texttt{LaMa} (\citealp{lama}) to implement direct numerical maximum likelihood methods that are compatible with automatic differentiation. 



\subsection{Implementation details: simulation}
In this section, we provide details on the simulated covariates used in scenarios 1 and 2 of the simulation study. 
The simulated covariate was generated as a deterministic time series defined as a combination of sine and cosine functions with a constant offset. Specifically, for $t = 1, \ldots, T$, we define
$$
u_t = 20 + 10\left\{ \sin\!\left(\frac{t}{150}\right) + \cos\!\left(\frac{t}{650}\right) \right\}.
$$

The amplitudes and frequencies were chosen to control the number of threshold crossings, and hence the proportion of observations assigned to the disturbed regime. The covariate sequence $\{u_t\}_{t=1}^T$ is deterministic and fixed across simulated datasets for a given sample size.

\begin{figure}[htbp]
    \centering
   \includegraphics[width=17cm]{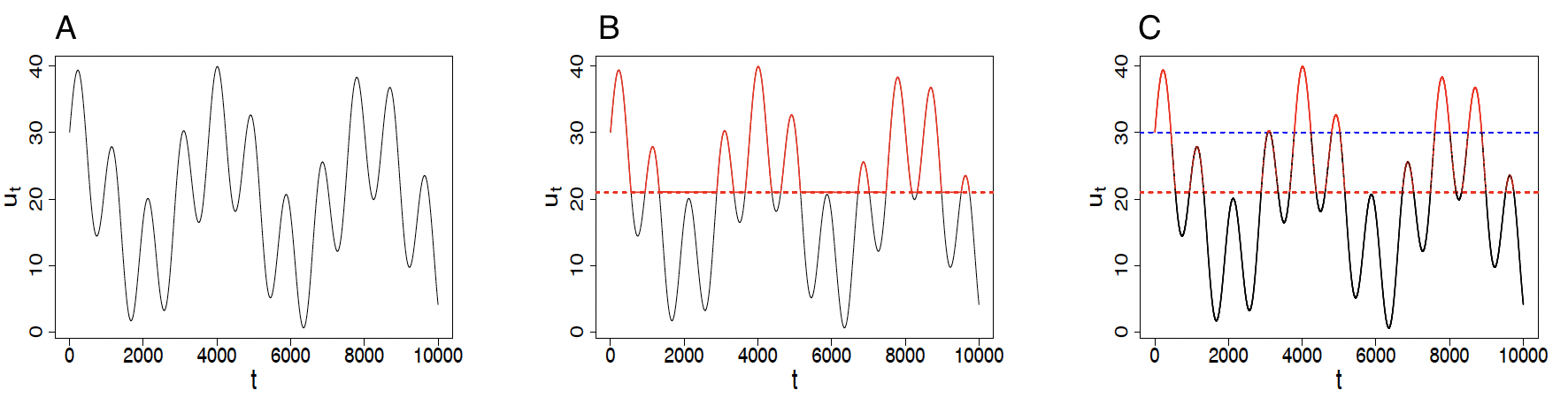}
    \caption{Time series of unstandardised covariates $\{{u}_{1,t}\}_{t=1}^T$ with the threshold values used in the simulation study: (a) corresponds to no disturbance (scenarios 1$.$b), (b) shows $\{{u}_{1,t}\}_t$ highlighted in red when exceeding the threshold defined in scenario 1$.$a, 
    and panel (c) shows $\{{u}_{1,t}\}_t$ in red when exceeding thresholds defined in scenario 2$.$a.}
    \label{fig:all_panels}
\end{figure}

Observe that, as shown in Figure \ref{fig:all_panels}c, in scenario 2$.$a the covariate is not always considered disturbed when exceeding 21. This occurs because we define
$$
\boldsymbol{u}_{3,t} =
\begin{cases}
(u_{1,t}, 0), & \text{if } u_{2,t} = 1, \\
(0, u_{1,t}), & \text{if } u_{2,t} = 0,
\end{cases}
$$
so that $u_{1,t}$ enters different dimensions depending on the value of $u_{2,t}$. Each dimension is associated with a different threshold ($21$ or $30$). Consequently, values of $u_{1,t}$ exceeding 21 may not be classified as disturbed when the relevant threshold is $30$, even though they would be if the indicator variable took the opposite value.

\subsection{Additional results}
\subsubsection{Description of bootstrap likelihood ratio test}
The bootstrap procedure involves generating $B = 100$ datasets based on the parameters of the null model (e.g., standard HMM) fitted to the data. For each bootstrap dataset, we fit both the null and alternative models and calculate the likelihood ratio $ 2(\ell_{H_1} - \ell_{H_0}) $, where $\ell_{H_1}$ refers to the likelihood of a THMM with two components and $\ell_{H_0}$ to the likelihood of a THMM under the null model. The empirical distribution of these bootstrap statistics is used to compute the p-value = $ b / B $, where $ b $ is the number of bootstrap likelihood ratios greater than the observed likelihood ratio. This approach corresponds to the standard parametric bootstrap likelihood ratio test (\cite{mclachlan_bootstrapping_1987, mclachlan2000finite, dziak_effect_2014}). The intuition is that when there is no measurable disturbance (i.e., under the null hypothesis), the observed data should closely align with the bootstrap datasets. Otherwise, they will be significantly different from each other. 
\subsubsection{Simulation Study}
\begin{figure}[H]
\hspace{2cm}
    \includegraphics[width=12cm,height=5.5cm]{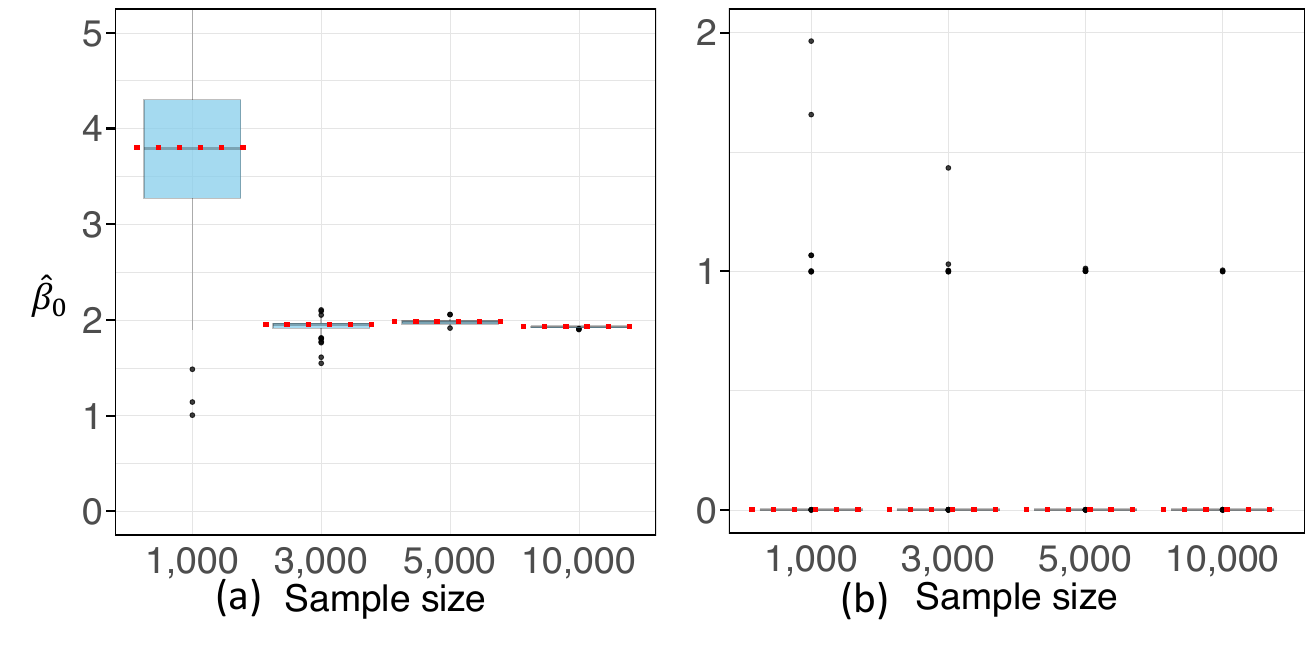}
    \caption[Estimates of ${\beta}_0$ obtained using the lasso-penalised THMM across different sample sizes and disturbance scenarios]{Estimates of ${\beta}_0$ obtained using the lasso-penalised THMM across different sample sizes (a) in the presence of disturbances (scenario 1$.$a) and (b) under the null model (scenario 1$.$b). The red dotted lines correspond to the true value of ${{\beta}}_0$ for different sample sizes. To improve readability, four outliers (estimates exceeding 2) from the sample size of $1,000$ were excluded in (b).}

    \label{resultsold1}
\end{figure}
\begin{figure}[H]
    \centering
    \resizebox{\textwidth}{!}{%
        \includegraphics{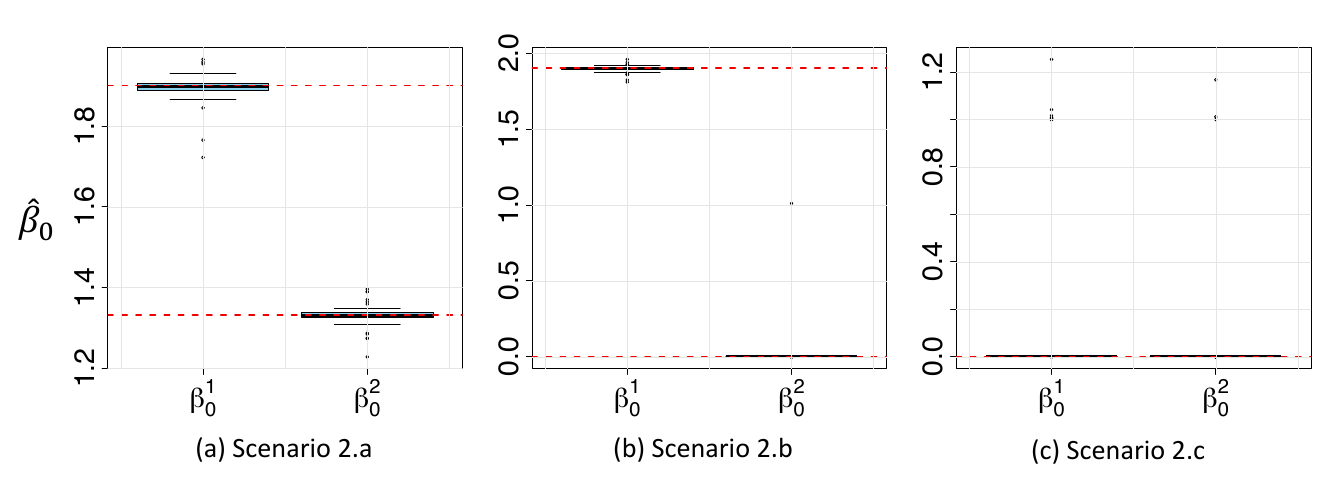}
    }
    \caption[Estimates of ${\boldsymbol{{\beta}}_0}$ obtained using the lasso-penalised THMM in the bivariate setting across disturbance scenarios]{Estimates of ${\boldsymbol{{\beta}}_0} = ({{\beta}}_0^1,{{\beta}}_0^2)$ obtained using the lasso-penalised THMM in the bivariate setting across different scenarios with sample size $10{,}000$: (a) both covariates have different disturbance thresholds $\boldsymbol{\beta}_0 = (1.90,1.33)$, (b) only one covariate has an active threshold $\boldsymbol{\beta}_0 = (1.90,0)$, and (c) neither covariate has a threshold $\boldsymbol{{\beta}}_0 = (0,0)$. The red dotted lines correspond to the true value for each element of ${\boldsymbol{{\beta}}_0}$.}
    \label{resultsold2}
\end{figure}

\begin{table}[h!]
\caption{Bias of penalised THMM estimates under different sample sizes and disturbance conditions}
\label{bias}
\hspace{-1.5cm}
\begin{tabular}{llcccccccc}
\toprule
 & & \multicolumn{8}{c}{Sample size} \\
\cmidrule(lr){3-10}
 & & \multicolumn{4}{c}{No Disturbance} & \multicolumn{4}{c}{Disturbance} \\
\cmidrule(lr){3-6} \cmidrule(lr){7-10}
 & Parameter (true value) & 1{,}000 & 3{,}000 & 5{,}000 & 10{,}000 & 1{,}000 & 3{,}000 & 5{,}000 & 10{,}000 \\
\midrule
\multirow{6}{*}{Bias} 
 & $\mu_0^1$ (1)   & -0.010 & 0     & 0.002 & 0     & 0.009  & 0     & 0.002 & -0.002 \\
 & $\mu_0^2$ (4)   & -0.006 & 0.01  & 0.004 & 0     & 0.109  & 0.02  & 0.015 & 0.005  \\
 & $\mu_0^3$ (10)  & 0.053  & 0.01  & 0.001 & 0.005 & -0.023 & -0.012 & 0.012 & 0      \\
 & $s_0^1$ (1.5)   & -0.001 & -0.007 & -0.01 & -0.003 & -0.023 & -0.012 & -0.011 & -0.04  \\
 & $s_0^2$ (10)    & 0.21   & 0.079 & 0.087 & 0.006 & 0.17   & 0.04  & 0.032 & 0.038  \\
 & $s_0^3$ (12)    & 0.26   & 0.103 & 0.047 & 0.006 & 0.23   & 0.17  & 0.069 & 0      \\
\bottomrule
\end{tabular}
\end{table}

\begin{table*}[ht]
\caption{False-positive rates, power, and computational costs (quartiles in minutes) for lasso and BLRT in scenarios 2$.$a–2$.$c with 
sample size
$10,000$.}
\label{tabtab}
\hspace{-2cm}
\begin{tabular}{lcccccc}
\toprule
& \multicolumn{3}{c}{Lasso} & \multicolumn{3}{c}{BLRT} \\
\cmidrule(lr){2-4} \cmidrule(lr){5-7}
 & Scenario 2.a & Scenario 2.b & Scenario 2.c & Scenario 2.a & Scenario 2.b & Scenario 2.c \\
\midrule
False positive rate & - & 0.02 & 0.06, 0.15 & - & 0.48 & 0.10, 0.03 \\   
Power ($\%$) & 100, 100 & 100 & - & 100 & 70 &  -\\  
Computational cost & 3-3.7 & 2.9-3.1 & 2.3-3.1 & 15.8-17.6 & 16.2-21.4 & 16.4-22.3 \\
\bottomrule
\end{tabular}
\end{table*}
\newpage\noindent Recall that scenarios 2$.$a--c all involve the covariate $$
\boldsymbol{u}_{3,t} =
\begin{cases}
(u_{1,t}, 0), & \text{if } u_{2,t} = 1, \\
(0, u_{1,t}), & \text{if } u_{2,t} = 0.
\end{cases}
$$ In scenario 2$.$a, both dimensions have disturbance effects with thresholds at $21$ and $30$ respectively, and only the full null model ($H_0: \boldsymbol{\beta_0} = (0,0)$) is used with the alternative model $ \boldsymbol{\beta_0} = (\beta^1_0 > 0,\beta^2_0 > 0)$.
In this scenario, the BLRT selects the alternative model over the null in 100\% of cases, correctly identifying the presence of a disturbance effect for every simulated dataset. The false-positive rate for the lasso is not defined in Scenario~2$.$a because both covariates have active thresholds, leaving no opportunity for false detections. However, as shown in Figure~3a of the main manuscript, the lasso consistently estimates both thresholds, with none shrunk to zero, corresponding to a power of 100\%.

In scenario 2$.$b, only the first dimension of the covariate $\{\boldsymbol{u}_{3,t}\}_{t=1}^T$ has a disturbance effect with a threshold at $21$, while the second has no disturbance effect (i.e., no threshold). In scenario 2$.$c, the covariate is not associated with any disturbance effect. For both scenarios 2$.$b and 2$.$c, all three null hypotheses are used with the bootstrap likelihood ratio test (BLRT):
$H_{01}\!: \boldsymbol{\beta}_0 = (0,0)$,
$H_{02}\!: \boldsymbol{\beta}_0 = (0,\beta_0^2 > 0)$, and
$H_{03}\!: \boldsymbol{\beta}_0 = (\beta_0^1 > 0, 0)$.
For readability, we present only the latter two in Table~\ref{tabtab}, while the remaining hypothesis is discussed in the text below. In scenario 2$.$b, the proportion of BLRT p-values below $0.05$ when using $H_{02}$ (i.e., $\beta_0^1=0$ and $\beta^2_0 > 0$) represents the Type I error (expected to be around $0.05$), while the proportion of p-values below $0.05$ when using $H_{03}$ represents the statistical power to detect the active threshold. In scenario 2$.$b, the BLRT always chooses the full model over the null, correctly detecting a disturbance in $100$\% of cases. However, it has difficulty identifying the active covariate, reaching only $70$\% power for detecting the threshold of covariate 1. Additionally, its Type I error rate is severe at about $48$\%. Moreover, only about half of the 50 simulated datasets converged when testing $H_{02}$ and $H_{03}$, thus these results are only based on 30 datasets. In scenario 2$.$c, the BLRT incorrectly favoured the alternative model over the full null 12\% of the time. However, its control of Type I error for each covariate dimension separately is better: $0.1$ for $\beta_0^1$ and $0.03$ for $\beta_0^2$, which could result in conflicting outcomes, where the full model is better than the null but no threshold is found significant. 

Note that none of the tests were corrected for multiple testing. In scenario 2$.$c, applying a Bonferroni correction for three tests alters the estimates, reducing the overall Type I error rate when testing against $H_{01}$ from $12$\% to $2$\%, while the specific Type I error rates remain unchanged. In scenario 2$.$b, applying the correction has little impact on the estimates, decreasing the Type I error rate when testing against $H_{02}$ from $48$\% to $40$\% and leaving the power unchanged.

\newpage

\begin{figure}[H]
    \centering
    \includegraphics[width=15.5cm,height=7.5cm]{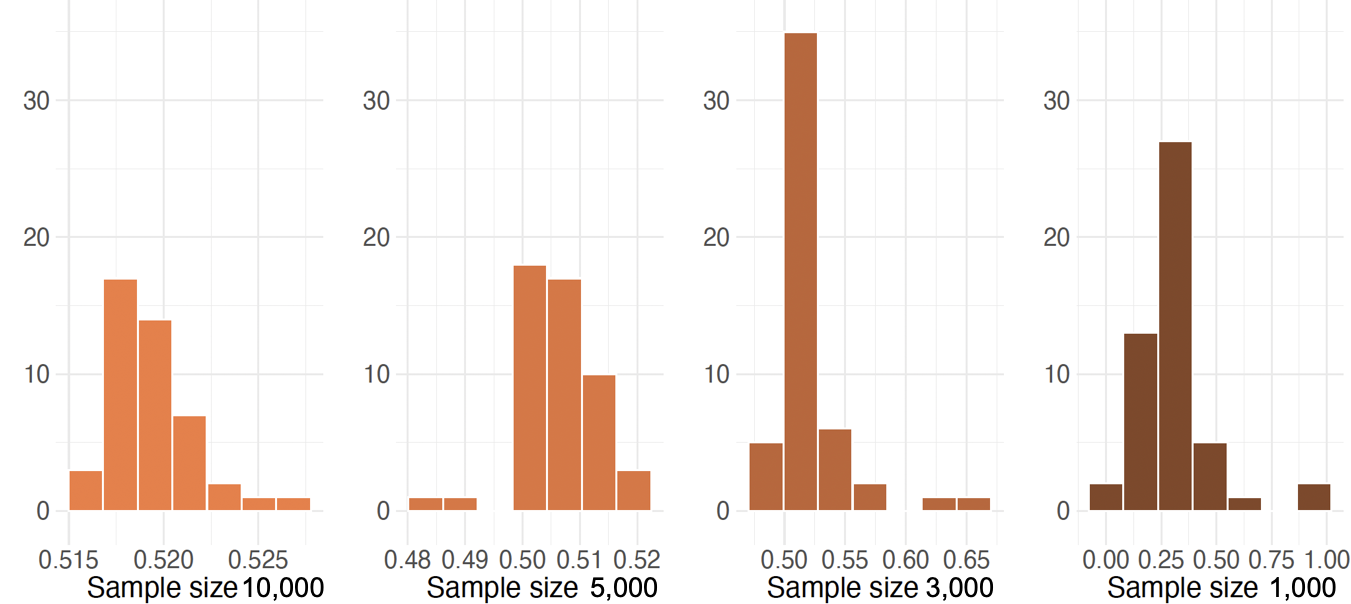}
    \caption{Histogram of $\hat{\lambda}$ across 50 datasets simulated in the presence of disturbance, in scenario 1$.$a.}
    \label{fig:lambda1}
\end{figure}

\begin{figure}[H]
    \centering
    \includegraphics[width=15cm,height=10cm]{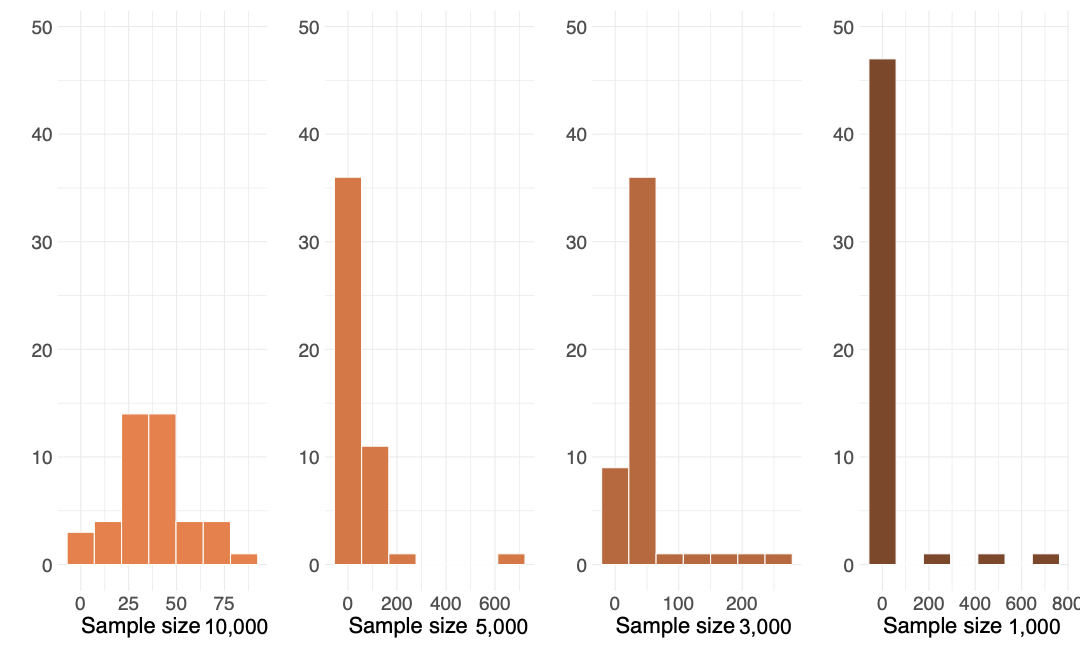}
    \caption{Histogram of $\hat{\lambda}$ across 50 datasets simulated under the null (no disturbance) in scenario 1$.$b, plotted on the log-scale.}
    \label{fig:lambda2}
\end{figure}

As discussed in the main manuscript for scenario~1, under disturbance (scenario~1$.$a), the distribution of $\hat{\lambda}$ is concentrated around its mean (Figure~\ref{fig:lambda1}). Under the null model (scenario~1$.$b), most $\hat{\lambda}$ values are very large, leading to strong shrinkage of $\hat{\beta}_0$ towards zero, although occasional outliers with $\hat{\lambda} < 1$ are observed.

\subsubsection{Narwhal movement data case study}

Recall that in the narwhal case study, we have $\boldsymbol{\beta}_0=(\beta_0^1, \beta_0^2)$, where $\beta_0^1$ captures the effect of exposure to vessel in the presence of land, and $\beta_0^2$ represents the effect of exposure in the absence of land. The model estimates are:

\[
\log\boldsymbol{\beta}_0 = -9.651577,   2.690258,
\]
which, after rescaling and inversion, correspond to distance thresholds of
\[
0\text{ km} \;\;\text{and}\;\; 3.41 \text{ km}.
\]

\paragraph*{Movement and depth parameters by state}
\[
\begin{array}{lccc}
\toprule
 & \text{State 1} & \text{State 2} & \text{State 3} \\
\midrule
\mu_\text{step}           & 0.99 & 2.57 & 1.32 \\
\text{shape}_\text{step}        & 2.23  & 13 & 3 \\
\kappa_{angle}      & 0.65 & 6.67 & 1.25 \\
\mu_{\max \mathrm{depth}} & 37.24 & 43.85 & 354.06 \\
\text{shape}_{\max \mathrm{depth}} & 1.29 & 1.18 & 5.17 \\
\bottomrule
\end{array}
\]

Based on the Viterbi algorithm, the estimated time allocation to each state is 31\% for State 1, 36\% for State 2, and 33\% for State 3.

\subsubsection{Narwhal movement data case study: comparative analysis}
In this subsection, we perform a comparative analysis to choose the baseline cutoff. The cutoff used in the main analysis was $77$ km, corresponding to the $60$th percentile of the distribution of distances to the nearest vessel. We additionally explored thresholds of $50$, $60$, $70$, and $100$ km; replication code is provided on github.

After rescaling and inversion, the estimated distance thresholds under each cutoff are as follows:
\begin{table}[h]
\centering
\begin{tabular}{lccccc}
\hline
 & \textbf{50 km} & \textbf{60 km} & \textbf{70 km} & \textbf{77 km} & \textbf{100 km} \\
\hline
Threshold 1 (km) & 0 & 0 & 0 & 0 & 0 \\
Threshold 2 (km)  & 3.41 & 3.60 & 3.41 & 3.41 & 0.57 \\
nllk             & 51613.43 & 51618.17 & 51613.43 & 51613.43 & 51619.24 \\
\hline
\end{tabular}
\caption{Estimated distance thresholds and negative log-likelihood (nllk) values under each candidate baseline cutoff.}
\label{tab:cutoff_thresholds}
\end{table}

Selecting the threshold that minimises the negative log-likelihood therefore recovers the $77$ km cutoff used in the main analysis, supporting its choice. The results are largely consistent across the $50$, $60$, and $70$ km thresholds. The slight discrepancy at 
$60$ km is likely attributable to an insufficient number of random initial values. The $100$ km case, however, yields a very different threshold associated with a higher negative log-likelihood. We believe this is due to the larger baseline cutoff assigning fewer observations to the baseline region, forcing the model to detect disturbance within a larger pool of data that is predominantly baseline behaviour. Since disturbed observations are rare, we believe their signal becomes difficult to isolate when overwhelmed by the volume of baseline observations. This phenomenon is examined in greater detail in the following subsection. 

\subsection{Additional simulations}
In this scenario, we simulate data with a single covariate associated with disturbance, using a threshold of $38$ (compared to $21$ in the main manuscript). For a sample size of $T = 10{,}000$, this corresponds to only $5\%$ of observations falling in the disturbed regime. The goal is to assess how well the method performs when disturbance events are rare.

As expected, the model's performance decreases when disturbed observations represent a small fraction of the data. However, it improves as the baseline cutoff increases (i.e., we allocate more data into the baseline component before model fitting): using the $0.10$, $0.90$, and $0.93$ quantiles of the covariate distribution as the baseline cutoff (i.e., $10\%$, $90\%$, and $93\%$ of the data, respectively, are allocated to baseline based on these quantiles), the method correctly detects disturbance $50\%$, $58\%$, and $67\%$ of the time, respectively. This pattern suggests that restricting the baseline to a larger portion of the data allows the disturbance signal to emerge more clearly. This finding is particularly relevant for the case study, where a baseline cutoff as large as $100$ km likely incorporates too much baseline data in the analysis, diluting the disturbance signal and rendering it difficult to detect.

\begin{figure}[h]
\includegraphics[width=15cm]{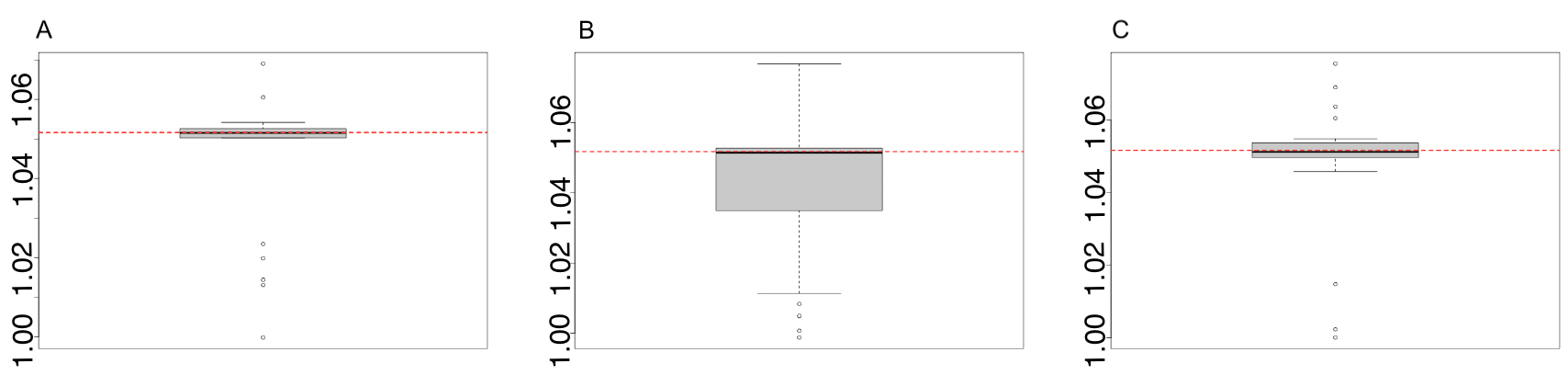}
    \caption{Boxplots of $\hat{\boldsymbol{\beta}}_0$ estimates across simulations for baseline cutoffs corresponding to the $0.10$ (A), $0.90$ (B), and $0.93$ (C) quantiles of the covariate distribution. The red dotted lines correspond to the true value of $\beta_0.$}
    \label{fig:boxplots_beta_quantiles}
\end{figure}
\subsection{Computational costs}
Below we report the computational costs of both methods, BLRT and lasso, across different scenarios.
\begin{figure}[htbp]
  \hspace{-1cm}
  \includegraphics[width=7.5cm,height=5cm]{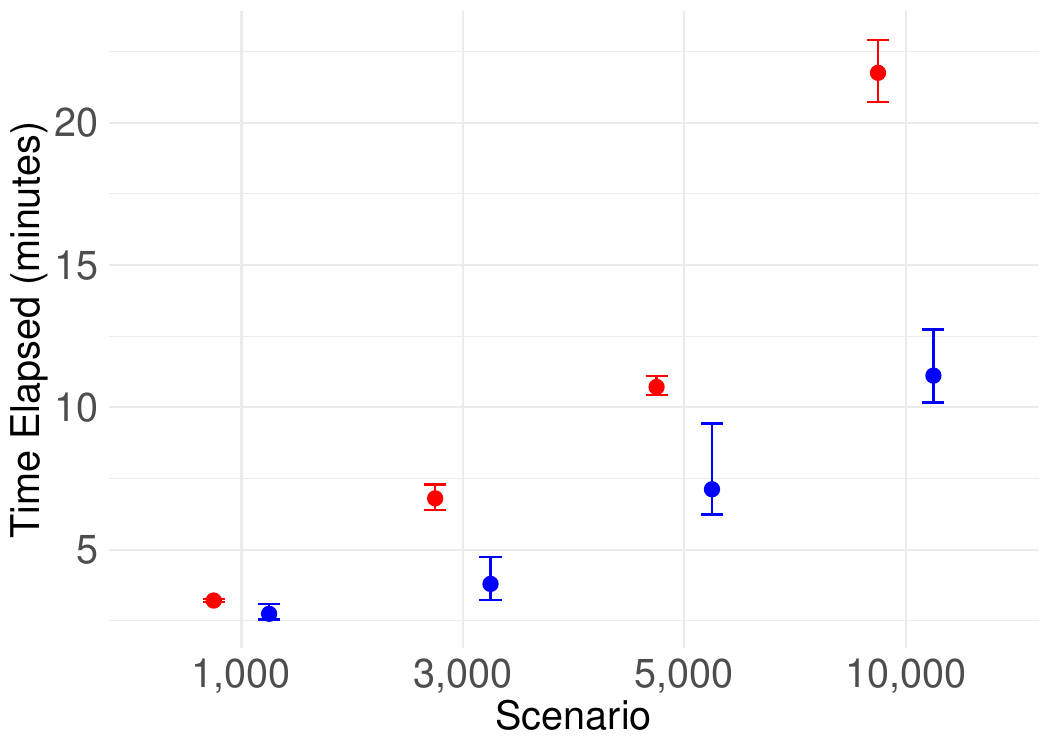}\hfill
  \includegraphics[width=7.5cm,height=5cm]{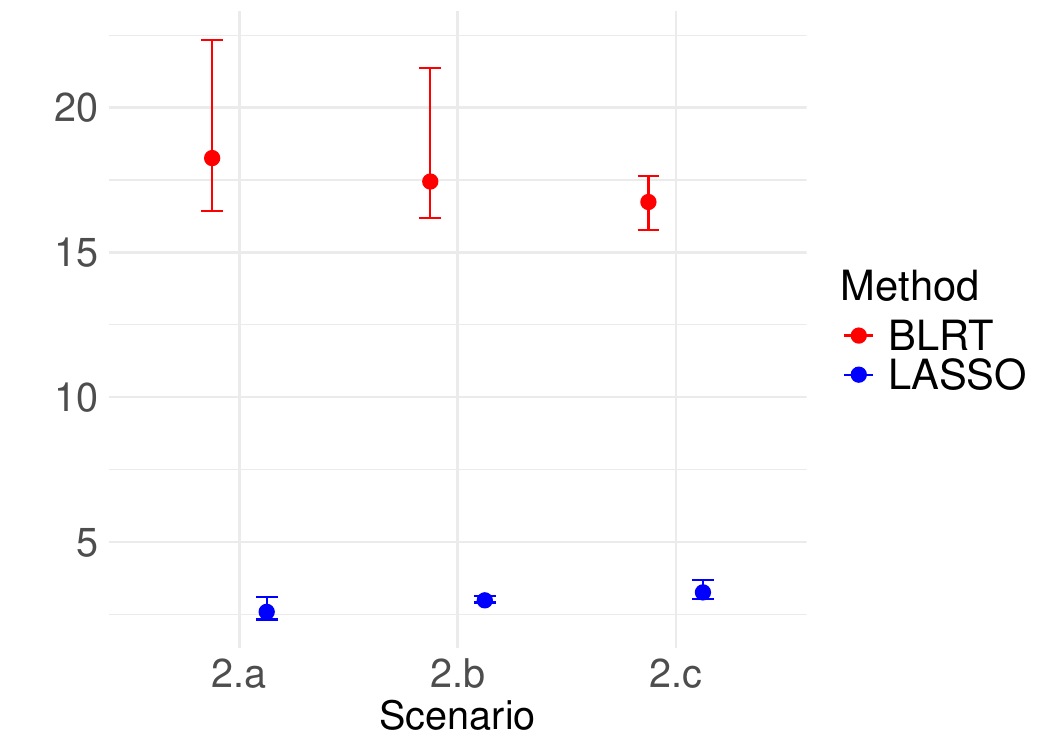}
  \caption{Computational costs for BLRT and THMM (lasso) for data simulated under the null model across different sample sizes for scenario 1$.$b (left) and for bivariate covariate scenarios, with sample size $10,000$ (right).}
  \label{fig:costplots}
\end{figure}
\newpage 
\subsection{Bias in qREML approach}
In this simulation, we justify the removal of the prior for $\boldsymbol{\theta}$ in equation (8) by showing that parameter estimates under a normal prior with precision $\epsilon$ converge to those from a model with no prior (i.e., omitting it) as $\epsilon \to 0$.

We simulated data from a THMM with a sample size of $10,000$, three states, and two hidden components (baseline and disturbed). For each dataset, we fitted seven different prior distributions, treating all model parameters as random effects. Specifically, the THMM parameters (excluding $\beta_0$) were assigned independent normal distribution with a shared precision $\epsilon$ (i.e., variance $= 1/\epsilon$), which was fixed and not estimated. We explored a range of $\epsilon$ values: $0.1$, $1$e-2, $1$e-3, $1$e-4, $1$e-5, and $1$e-6. For each value of $\epsilon$, we fitted the THMM, and selected the lasso regularisation parameter using the Laplace approximation combined with qREML. The results are based on approximately $40$ simulated datasets and indicate minimal bias for $\epsilon < 1$e-2 for all parameters (dotted line represents the true parameter value).

The simulation results show that the bias in the parameter estimates approaches zero as the prior precision $\epsilon$ tends to zero. This confirms that using a normal prior with very low precision is asymptotically equivalent to omitting the prior from the derivation, as done in equation (8). The only theoretical concern is that, as $\epsilon$ approaches zero, the normal distribution becomes improper. However, this has not posed practical problems for estimation in the simulation and case studies.

\begin{figure}[htbp]
  \centering
  \includegraphics[width=\linewidth]{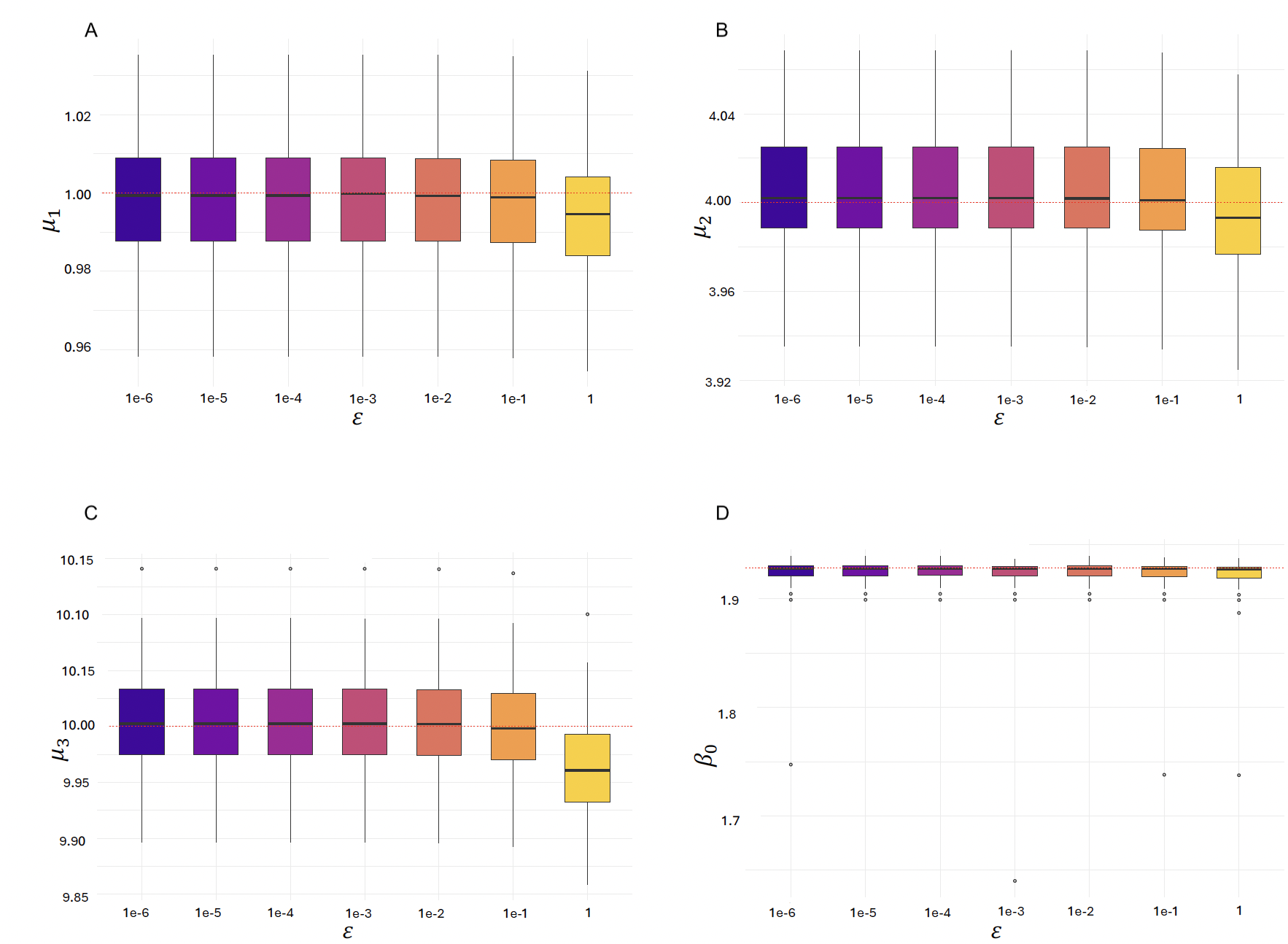}
  \caption{Parameter estimates when the normal prior is included in the qREML, with different precision values ($\epsilon$)}
  \label{fig:mu_beta}
\end{figure}

\end{document}